\documentclass[twocolumn,showpacs,preprintnumbers,amsmath,amssymb]{revtex4-1}
\usepackage{ragged2e}
\usepackage{graphicx}
\usepackage{dcolumn}
\usepackage{bm}
\usepackage{ulem}
\usepackage[]{color}
\newcommand{\red}[1]{\textcolor{red}{#1}}

\begin{document}

\preprint{APS/123-QED}

\title{Antiferromagnetic correlations in strongly valence fluctuating CeIrSn}

\author{Y. Shimura$^{1}$}
\thanks{These authors contributed equally to this work}
\author{A. W\"orl$^{2}$}
\thanks{These authors contributed equally to this work}
\author{M.~Sundermann$^{3,4}$}
\thanks{These authors contributed equally to this work}
\author{S. Tsuda$^{1}$}
\author{D. T. Adroja$^{5,6}$}
\email{devashibhai.adroja@stfc.ac.uk}
\author{A. Bhattacharyya$^{7}$}
\author{A. M. Strydom$^{6}$} 
\author{A. D. Hillier$^{5}$} 
\author{F. Pratt$^{5}$}
\author{A.~Gloskovskii$^{4}$}
\author{A.~Severing$^{3,8}$}
\author{T. Onimaru$^{1}$}
\author{P. Gegenwart$^{2}$}
\email{philipp.gegenwart@physik.uni-augsburg.de}
\author{T. Takabatake$^{1}$}

\affiliation{
$^{1}$ Graduate School of Advanced Science and Engineering, Hiroshima University, Higashi-Hiroshima, 739-8530, Japan \\
$^{2}$ Experimental Physics VI, Center for Electronic Correlations and Magnetism, University of Augsburg, 86159 Augsburg, Germany  \\
$^{3}$ Max Planck Institute for Chemical Physics of Solids, 01187 Dresden, Germany \\
$^{4}$ Deutsches Elektronen-Synchrotron DESY, 22607 Hamburg, Germany \\
$^{5}$ ISIS Facility, Rutherford Appleton Laboratory, Chilton, Didcot Oxon, OX11 0QX, United Kingdom \\
$^{6}$ Highly Correlated Matter Research Group, Physics Department, University of Johannesburg, PO Box 524, Auckland Park 2006, South Africa \\
$^{7}$ Department of Physics, Ramakrishna Mission Vivekananda Educational and Research Institute, Belur Math, Howrah 711202, West Bengal, India \\
$^{8}$ Institute of Physics II, University of Cologne, 50937 Cologne, Germany
}

\begin{abstract}
CeIrSn with a quasikagome Ce lattice in the hexagonal basal plane is a strongly valence fluctuating compound, as we confirm by hard x-ray photoelectron spectroscopy and inelastic neutron scattering, with a high Kondo temperature of $T_{\mathrm{K}}\sim 480$\,K.  We report a negative in-plane thermal expansion $\alpha/T$ below 2\,K, which passes through a broad minimum near 0.75\,K. Volume and $a$-axis magnetostriction for $B \parallel a$ are markedly negative at low fields and change sign before a sharp metamagnetic anomaly at 6\,T. These behaviors are unexpected for Ce-based intermediate valence systems, which should feature positive expansivity. Rather they point towards antiferromagnetic correlations at very low temperatures. This is supported by muon spin relaxation measurements down to 0.1\,K, which provide microscopic evidence for a broad distribution of internal magnetic fields. Comparison with isostructural CeRhSn suggests that these antiferromagnetic correlations emerging at $T\ll T_{\mathrm{K}}$ result from geometrical frustration.
\end{abstract}

\maketitle

Heavy fermion behavior in Ce- and Yb-based intermetallic compounds arises from the hybridization of localized 4f moments with conduction electrons. The Kondo interaction increases exponentially with the antiferromagnetic (AF) exchange $J$ between the local 4f and conduction electron spins, $T_{\rm K}~\sim \exp(-1/J)$, while the indirect exchange coupling between the moments varies like $T_{\rm RKKY} \sim J^2$. For large $J$, $T_{\rm K}$ dominates and a paramagnetic Fermi liquid ground state is expected with all 4f moments bound in non-magnetic singlets. For very large $T_{\rm K}$, exceeding $\sim 10^2$\,K, not only the spin but also the 4f charge fluctuates, giving rise to intermediate valence behavior~\cite{ValenceBook}.This
 directly manifests itself in the volume, which undergoes characteristic changes as a function of temperature or magnetic field. Since non-magnetic Ce$^{4+}$ has a smaller ionic radius than magnetic Ce$^{3+}$, intermediate valence behavior results in positive thermal expansion, as in ordinary metals, and a positive magnetostriction due to the stabilization of the Ce$^{3+}$ valence state by magnetic field~\cite{Valence_Hafner85,CeSn3_Edelstein83}. In this Letter, we report a {\it negative} low-temperature thermal expansion and magnetostriction in the intermediate valent CeIrSn and provide microscopic evidence for AF correlations developing at temperatures two orders of magnitude below the large $T_{\rm K} \sim 480$\,K. 

CeIrSn and its sister compound CeRhSn crystallize in the hexagonal ZrNiAl-type structure in which the Ce atoms form a quasikagome lattice in the basal plane \cite{ZrNiAl_Hulliger93,CeTSn_Chevalier06,ZrNiAl_Pottgen15}.
The Kondo temperatures were estimated as $T_{\rm K}\sim  240$\,K~(Rh) and $\sim 480$\,K~(Ir), from the Kondo maximum found in electrical resistivity and thermopower \cite{CeTSn_Bando00}.
These high Kondo temperatures classify the materials as valence fluctuating. 
Nevertheless, both compounds exhibit non-Fermi liquid behavior below $\sim$ 2\,K \cite{CeRhSn_Kim13, CeIrSn_Tsuda18}. 
For CeRhSn, AF spin fluctuations were detected by nuclear magnetic resonance measurements
over a wide temperature range between 1.3 K and 200 K ~\cite{CeRhSn_Tou04},
 while down to 0.05 K, muon spin rotation ($\mu$SR) experiments confirmed the absence of long-range magnetic order  \cite{CeRhSn_Schenck04}.
The magnetization shows a strong Ising-type anisotropy demonstrated by the magnetic susceptibility $\chi $ which exhibits $\chi {\rm _c} / \chi _{\rm a} \sim 10$.
CeRhSn and CeIrSn exhibit anisotropic metamagnetic crossovers only for fields applied perpendicular to the (easy) $c$-axis
 at $B_{\rm M} =$ 3.5\,T below 0.5\,K and 5.5\,T below 2\,K, respectively \cite{CeRhSn_Yang17, CeRhSn_Tokiwa15, CeIrSn_Tsuda18}.
For metamagnetic crossovers in ordinary Kondo lattice systems, the energy scale of $\mu _{\rm B} B_{\rm M}$ should be comparable to $k_{\rm B} T_{\rm K}$, where $\mu _{\rm B}$ and $k_{\rm B}$ are the Bohr magneton and the Boltzmann constant, respectively \cite{YbIr2Zn20_Takeuchi10}.
Strikingly, the $B_{\rm M}$ values of both materials are two orders of magnitude smaller than expected from the Kondo/valence fluctuation energy scales of 240\,K (Rh) and 480\,K (Ir). 

The emergence of metamagnetic signatures deep inside the valence fluctuating regime implies magnetic correlations. Thermal expansion and magnetostriction have proven to be highly suitable thermodynamic probes to investigate this behavior. Since the expansion coefficient measures the pressure-derivative of entropy, the entropy accumulation near a quantum critical point (QCP) results in a sign change of thermal expansion~\cite{QCP_Garst05}. In particular, a characteristically negative thermal expansion is observed when the ground state is tuned through a magnetic QCP to an AF ordered state \cite{CeRhIn5_Donath09, CeCu6_Grube18, CeSn3_Kuchler06}. 

For CeRhSn, the linear thermal expansion coefficient along the $a$-axis divided by temperature $\alpha _{\rm a}/T$ divergently increases on cooling, while Fermi liquid behavior is realized along the $c$-axis \cite{CeRhSn_Tokiwa15}. This indicates that the ground state is located in the vicinity of a QCP which is sensitive to uniaxial stress along the $a$-axis but insensitive to $c$-axis stress. Indeed, a sign change and anomaly of $\alpha _{\rm a}/T$ indicative of long-range magnetic order have been found at \red{$\sim$} 0.4\,K for $a$-axis uniaxial pressure $p_{\mathrm{a}}\ge 0.06$\,GPa~\cite{CeRhSn_Kuchler17}. The anisotropic quantum criticality and the formation of magnetic order for in-plane stress have been associated with the Kagome structure, whose frustration can be weakened by in-plane distortion. More direct evidence for geometrical frustration has been reported for the isostructural compound CePdAl, which displays partial magnetic order below $T_{\rm N} = 2.7$\,K \cite{CePdAl_Oyamada08, CePdAl_Donni96} with only 2/3 of the Ce moments being involved down to 0.03\,K.
In the isostructual CePtPb, related residual spin dynamics inside the AF ordered phase has been detected by $\mu$SR measurements \cite{CePtPb_Fang19}.
In this Letter, we prove that isostructural CeIrSn, despite being a strongly valence fluctuating Kondo lattice, features bulk AF correlations at temperatures below 2\,K.

\begin{figure}[t]
	\begin{center}
		\includegraphics[width=65mm]{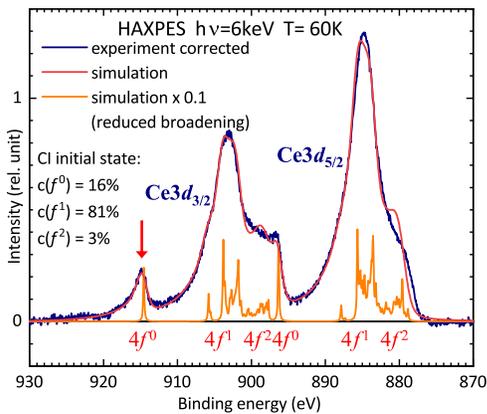}
	\end{center}
	\caption{Ce $3d$ core-level spectrum. Blue curve: as extracted from hard x-ray photoelectron spectroscopy (HAXPES); red curve: full-multiplet configuration-interaction simulation; orange curve: simulation with reduced broadening. The weights of the $4f^0$, $4f^1$, and $4f^2$ configurations in the ground state are also indicated.}
 \label{HAXPES}
\end{figure}

The strongly valence fluctuating ground state of CeIrSn is confirmed by HAXPES. For details on set-up, data treatment, and analysis see supplemental material (SM) \cite{SM} \nocite{Pratt00, CeRhGe3, CePd3_Murani96, Jones91, Stephenson88, Maisuradze18}  and also Refs.\,\cite{Strigari2015,Sundermann2016,Sundermann2017}. Figure\,\ref{HAXPES} displays the Ce3$d$ core-level emission spectrum of CeIrSn (blue line). The data show the typical 4$f^0$, 4$f^1$, and 4$f^2$ derived structures often observed in Ce intermetallic compounds. It is remarkable that the 4$f^0$ spectral weight at 915\,eV (see red arrow) is almost as strong as in the $\alpha$-type cerium compound CePd$_3$\,\cite{Sundermann2016}. A combined full-multiplet and configuration-interaction analysis\,\cite{Imer1987,Tanaka1994} provides an excellent simulation of the experimental spectrum. The red line displays the result taking into account resolution and lifetime broadening as well as some Mahan asymmetry (SM). The orange line shows the simulation with reduced broadening to better reveal the underlying multiplet structure of the three atomic configurations. The analysis indicates that the Ce ground state contains as much as 16\% 4$f^0$ configuration, thus, demonstrates unambiguously the intermediate valent nature of CeIrSn. We like to point out that the Ce3$p$ core level spectrum can be described with the same parameters (SM). Inelastic neutron scattering (INS) reveals absence of crystal-electric field excitations (SM), consistent with strong valence fluctuations evidenced by HAXPES.


\begin{figure}[t]
	\begin{center}
		\includegraphics[width=75mm]{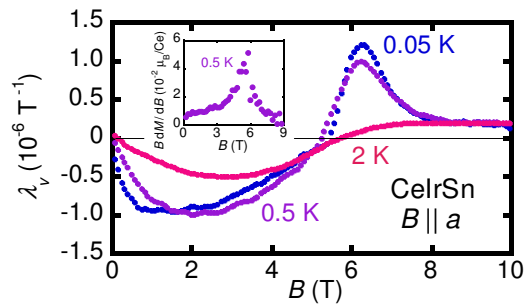}
	\end{center}
	\caption{Volume magnetostriction coefficient $\lambda _V=V^{-1}dV/dB$ for $B$ $||$ $a$, determined from three independent linear magnetostriction measurements (SM). The inset shows the field variation of $B \, {\rm d}M/{\rm d}B$ for $B$ $||$ $a$ at 0.5 K, calculated from the DC magnetization data extracted from Ref. \cite{CeIrSn_Tsuda18}.}
\end{figure}

Figure 2 displays the volume magnetostriction coefficient $\lambda _V=V^{-1}dV/dB$ for magnetic field along the $a$-axis at low temperatures, determined by high-resolution capacitive dilatometry~\cite{Dilatometer_Kuchler17} along three orthogonal directions (SM). At 0.05\,K and 0.5\,K, it exhibits a clear peak at $B_{\rm M} = $ 6\,T which agrees with the metamagnetic crossover fields determined by DC magnetization $M(B)$ and AC susceptibility $\chi _{\rm AC}(B)$ \cite{CeIrSn_Tsuda18}.
At 2\,K, the peak is suppressed and only a broad shoulder remains. Interestingly, for fields below $B_{\rm M}$, a negative $\lambda _V$ is found which becomes more pronounced as temperature decreases. The Maxwell relation
$\left( \frac{\partial V}{\partial B} \right)_P = - \left( \frac{\partial M}{\partial P} \right)_V$
indicates that the magnetization exhibits a positive hydrostatic pressure dependence below $B_{\rm M}$.
This finding is in stark contrast to conventional Ce-based non-magnetic valence fluctuating compounds,
 for which the application of hydrostatic pressure stabilizes the non-magnetic Ce$^{4+}$ state and therefore suppresses the magnetization \cite{Valence_Hafner85}.
As shown in the SM, the negative $\lambda _V$ mainly arises from the largely negative longitudinal linear magnetostriction $\lambda _a$ which indicates that uniaxial pressure along the $a$-axis increases the magnetization for fields below 5\,T. 

%

The sign change of $\lambda _V$ from negative to positive in the vicinity of $B_{\rm M}$ found in CeIrSn is unconventional among
 valence fluctuating and Kondo lattice systems with a metamagnetic crossover. A conventional example is CeRu$_2$Si$_2$ which exhibits a sharp metamagnetic crossover at $B_{\rm M}$ = 7.7\,T \cite{CeRu2Si2_Haen87}, where $\lambda _V$ is positive below and above $B_{\rm M}$~\cite{CeRu2Si2_Lacerda89, CeRu2Si2_Visser91}. Furthermore, the field dependence of $\lambda _V$ is related to the differential susceptibility $\chi(B)={\rm d}M(B)/{\rm d}B$ by a single constant positive parameter $\Omega $ \cite{CeRu2Si2_Puech88, CeRu2Si2_Matsuhira99} as
\begin{equation}
\lambda _V (B) = \Omega B \chi.
\end{equation}
The parameter $\Omega = -{\rm d}\log B_{\rm M}/{\rm d}\log V = 1/\kappa_T (B_{\rm M})^{-1}({\rm d}B_{\rm M}/{\rm d}p)$
 ($\kappa_T$: isothermal compressibility) quantifies the volume dependence of the metamagnetic transition field \cite{Gruneisen_Kaiser88}.
It is constant if the free energy scales with respect to $B/B_{\rm M}$, implying that $B_{\rm M}$ is the dominating field scale. For Ce-based Kondo systems, the Kondo scale and thus also the metamagnetic transition field increase with pressure, giving rise to a positive scaling parameter $\Omega$ and, since the other factors in Eq. (1) are also positive, $\lambda_V>0$. 
In contrast to Ce systems, the Kondo scale of Yb systems decreases under pressure and thus $\lambda_V < 0$ is expected. 
For instance, in YbCu$_5$ ($B_{\rm M} = 17 $ T) and YbIr$_2$Zn$_{20}$ ($B_{\rm M} = 9.7 $ T for $B$ $||$ [100]) $\lambda _V$ remains negative below and above $B_{\rm M}$ and follows the magnetic susceptibility \cite{YbCu5_Tsujii01, YbIr2Zn20_Takeuchi10, YbIr2Zn20_Shimura10}. Generally, Eq. (1)
 requires that one dominating parameter governs the metamagnetic behavior.

The inset of Fig.~2 displays $B {\rm d}M/{\rm d}B$ as a function of $B$ $||$ $a$ for CeIrSn at 0.5\,K calculated from the DC magnetization data in Ref. \cite{CeIrSn_Tsuda18},
 which is positive in the whole field range. 
Thus, the sign change of $\lambda_V$ below the metamagnetic transition indicates that the scaling is violated,
 in contrast to the conventional paramagnetic heavy-fermion systems described above.
Indeed, the negative sign of $\lambda_V$ at low fields is unusual for Ce-based systems
 and fully unexpected for an intermediate valence material with $T_{\rm K}$ as large as 480\,K. 
Since magnetostriction is a volume sensitive probe, there is no way to associate this anomalous behavior with spurious secondary or impurity phases.
The observed negative $\lambda_V$, through the Maxwell relation indicates an increase of the magnetization with pressure.
Such increase is typically associated with the suppression of AF order or AF correlations by pressure.
In fact, a similar negative-positive sign change of $\lambda_V$ at $B_{\rm M}$ was also observed
 in the Ce-based antiferromagnets CeAl$_2$ and CeRh$_2$Si$_2$ \cite{CeAl2_Eric91, CeRh2Si2_Naito03}.
Thus, the negative magnetostriction $\lambda_V < 0$ for CeIrSn suggests the development of cooperative magnetism at low fields and low temperatures. 
Even above $B_{\rm M}$, the scaling of Eq. (1) does not work. In fact, it is only expected when the field energy equals the valence fluctuation energy, which will require high fields above 100 T for CeIrSn. 
\begin{figure}[t]
	\begin{center}
		\includegraphics[width=80mm]{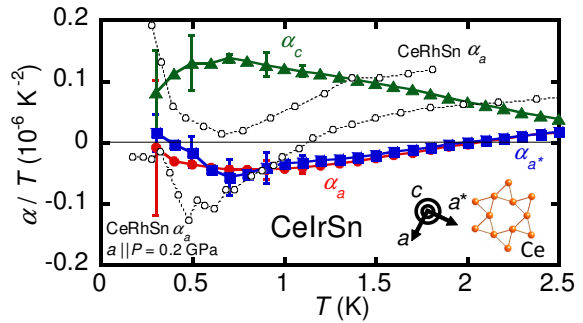}
	\end{center}
	\caption{Linear thermal-expansion coefficient as $\alpha /T$ of CeIrSn along the $a$ (red circles), $c$ (green triangles), and $a^*$ (blue squares) axes at zero field.
		$\alpha _a/T$ data for CeRhSn at almost zero pressure and at 0.2~GPa applied along the $a$ axis are extracted from Refs. \cite{CeRhSn_Tokiwa15, CeRhSn_Kuchler17}.
	}
\end{figure}

Figure 3 displays measurements of the linear thermal expansion $\alpha_i=L_i^{-1}dL_i/dT$ along the three main axes $i=a$, $a^*$ and $c$ (cf. the sketch) as $\alpha_i/T$ vs. $T$ at zero field. The in-plane expansivities are negative below 2\,K and show a minimum at $\sim 0.75$\,K, in contrast to a positive $\alpha _c/T$.
For comparison, we also plot $\alpha _a/T$ of CeRhSn at almost ambient pressure and under a uniaxial stress $P$ $||$ $a$ of $P = 0.2 $\,GPa \cite{CeRhSn_Tokiwa15, CeRhSn_Kuchler17}. 
Note that $\alpha /T$ of CeRhSn, measured at almost ambient pressure, diverges only along $a$, indicating that quantum criticality only couples to in-plane stress. This suggests quantum criticality driven by geometrical frustration and indeed the application of a distorting $a$-axis uniaxial stress releases the non-Fermi liquid behavior and leads to a sign change of $\alpha$ at low temperatures.

As shown in Fig. 3, the $\alpha /T$ data of CeIrSn along the three directions do not diverge upon cooling to zero. This finding is in stark contrast to the low temperature divergence in $\alpha _a/T$ of CeRhSn at almost ambient pressure
 but qualitatively resembles the behavior found at a uniaxial stress of 0.2\,GPa.
The uniaxial stress induced discontinuous change of $\alpha_a/T$ in CeRhSn is attributed to magnetic ordering \cite{CeRhSn_Kuchler17}. 
The negative in-plane thermal expansion without discontinuity observed in CeIrSn may thus indicate a short-range order. 
We recall that a negative thermal expansion manifests itself in Ce-based compounds
 such as CeCu$_{6-x}$Au$_x$ and CeRhIn$_{5-x}$Sn$_x$ when crossing the magnetic QCPs from a paramagnetic to an AF ordered state \cite{CeRhIn5_Donath09, CeCu6_Grube18}. 


The anomalous negative in-plane thermal expansion and negative volume magnetostriction suggest that, despite its huge Kondo scale of 480\,K, CeIrSn features a magnetic ground state. 
In order to obtain further microscopic information on the magnetic properties,
we utilized $\mu$SR measurements. 
Figure 4(a) shows the ZF-$\mu$SR spectra measured at various temperatures ranging from 0.1\,K to 4\,K. 
The ZF-$\mu$SR data show neither loss of the initial asymmetry nor any sign of frequency oscillations down to 0.1\,K. 
For a classical long-range magnetically ordered system below the magnetic transition temperature, ZF-$\mu$SR spectra exhibit either frequency oscillations
 for a small ordered moment or loss of initial muon asymmetry for a larger ordered moment
 (typically observed in ISIS-pulsed muon studies due to pulsed width of the muon beam). 
The ZF-$\mu$SR results indicate the presence of two components in the relaxation (Lorentzian form) below 2.5\,K,
 while the spectra can be fitted with one slow component above 2.5\,K. 

\begin{figure}[t]
	\begin{center}
		\includegraphics[width=70mm]{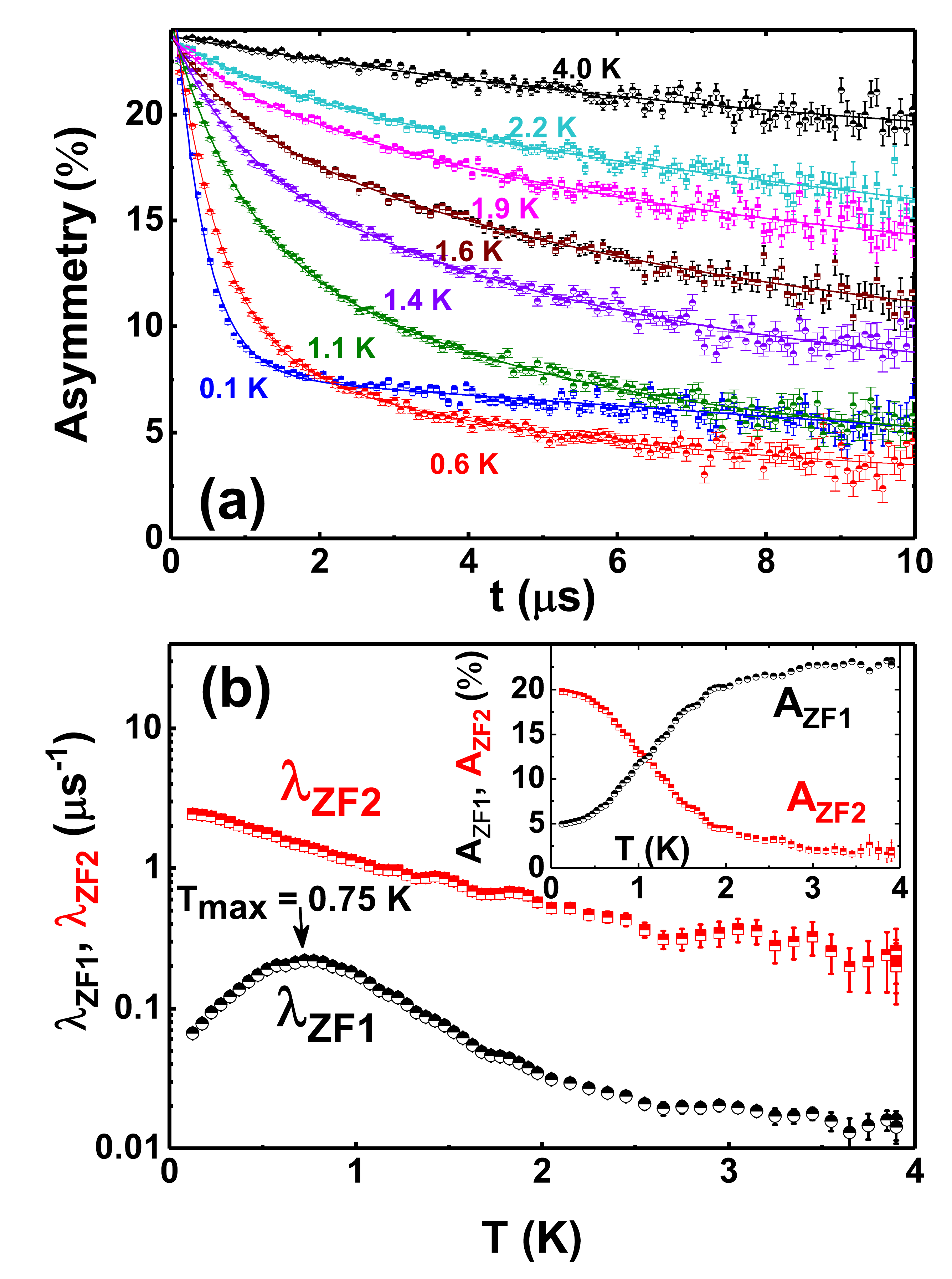}
	\end{center}
\centering
\caption{(a) Zero field $\mu$SR spectra collected at the indicated temperatures.
The solid curves are fits to the data by use of Eq.~(\ref{eq:ZF_MuSR}).
(b) Temperature dependence of the slow ($\lambda_{\mathrm{ZF1}}$) and the fast ($\lambda_{\mathrm{ZF2}}$) relaxations.
The inset shows the temperature dependence of the initial asymmetry of the slow relaxations ($A_{\mathrm{ZF1}}$) and the fast one ($A_{\mathrm{ZF2}}$).}
\end{figure}

The ZF-$\mu$SR spectra were fitted using various forms of the relaxation functions with one or two components. 
The best fits were obtained using Eq.~(\ref{eq:ZF_MuSR}) given below.

\begin{equation}
 G_z(t)= A_{\mathrm{ZF1}} e^{-\lambda_{\mathrm{ZF1}} t} + A_{\mathrm{ZF2}} e^{-\lambda_{\mathrm{ZF2}} t}+A_{\mathrm{BG}},
\label{eq:ZF_MuSR}
\end{equation}

where $A_{\mathrm{ZF1}}$ ($A_{\mathrm{ZF2}}$) and $\lambda_{\mathrm{ZF1}}$ ($\lambda_{\mathrm{ZF2}}$) are the initial asymmetries and the electronic relaxation rates of the slow (fast) components, respectively.
$A_{\mathrm{BG}}$ is the non-relaxing background contribution from the silver sample holder, which was estimated to be 5\,\% by fitting the 0.1\,K spectra and was kept fixed.
The results of the fit are presented in Fig. 4(b). 
At 4\,K, the initial asymmetry ($A_{\mathrm{ZF2}}$) of the fast relaxation is almost negligible (2\,$\%$), while the one of the slow relaxation ($A_\mathrm{ZF1}$) dominates (23\,$\%$). 
With decreasing temperature, $A_{\mathrm{ZF1}}$ starts decreasing while $A_{\mathrm{ZF2}}$ starts increasing. 
After they cross near 1.2\,K, the total asymmetry ($A_{\mathrm{ZF1}}$+$A_{\mathrm{ZF2}}$) remains temperature independent.
At 0.1\,K, $A_{\mathrm{ZF2}}$ is dominant (20\,$\%$) and $A_{\mathrm{ZF1}}$ has smaller (5\,$\%$) contribution. 
In contrast to the increase in the fast relaxation $\lambda_{\mathrm{ZF2}}$ on cooling, the slow relaxation component $\lambda_{\mathrm{ZF1}}$ exhibits a peak at $\sim $ 0.75 K,
 which indicates the development of dynamical short-range magnetic correlations. 
Notably, this temperature agrees well with the temperature where $\alpha _a/T$ shows a minimum.
The overall behavior of ZF-$\mu$SR spectra shows the presence of a very broad distribution of internal field without coherent precession of muon spins (see Fig. S6 in the SM).
The Lorentzian relaxation function fits the $\mu$SR data of CeIrSn which is better than the stretch-exponential or root exponential functions used for a spin-glass system.
The results of this analysis indicate that the ground state of CeIrSn is different from the classical spin-glass phase.
These results are supported through the distribution of a broad internal field in our TF-$\mu$SR presented in the SM. The ZF-$\mu$SR spectra of CeRhSn exhibit only one Lorentzian component~\cite{CeRhSn_Schenck04} with a weak anomaly near 1\,K, while the asymmetry exhibits a similar temperature dependence as the one of $A_{\mathrm{ZF2}}$ in CeIrSn. Furthermore, the temperature dependences of the ZF- and TF-$\mu$SR asymmetries in CeIrSn indicate homogeneous magnetism without any phase separation. Thus, CeIrSn has two relaxation components.
We note that a coexistence of magnetic order and magnetic fluctuating components was observed in the isostructural antiferromagnets CePdAl and CePtPb \cite{CePdAl_Oyamada08, CePdAl_Donni96, CePtPb_Fang19}, likely caused by the frustrated quasikagome lattice.
Since these Ce$TX$ systems including Ce(Pd$_x$Rh$_{1-x}$)Sn ($0.75 > x > 0.20$) with the ZrNiAl-type structure exhibit AF ordering,
 we conjecture that the magnetic correlations in CeIrSn are also of AF type \cite{CeRhSn_Yang17}.

As discussed above, application of $a$-axis uniaxial stress, which evades frustration, has induced low-temperature magnetic ordering in CeRhSn~\cite{CeRhSn_Kuchler17}. Such $a$-axis stressed CeRhSn and CeIrSn share similar features such as a negative in-plane linear thermal expansion due to magnetic correlations.
The negative magnetostriction along the $a$-axis in CeIrSn also indicates that the magnetism is highly sensitive to in-plane uniaxial stress.
Thus, the strong valence fluctuating material CeIrSn exhibits a highly uniaxial pressure-sensitive magnetism below 2\,K. 
Reflecting the magnetic correlations, the specific heat divided by temperature $C(T)/T$ increases on cooling below $\sim $ 4 K
 and tends to saturate around 2 K with a moderately large value of 80 mJ/K$^2$mol, in spite of the strong valence fluctuations \cite{CeIrSn_Tsuda18}.
This characteristic temperature $T^* \sim 2 $ K is two orders of magnitude smaller than $T_{\rm K} \sim 480 $ K.
Emergence of magnetic correlations has been reported in various heavy-fermion metamagnets without long-range ordering like CeRu$_2$Si$_2$,
for which, however, $T^* $ is close to $ T_{\rm K}$ \cite{metamag_Aoki13}.
The giant difference between $T^*$ and $T_{\rm K}$ in CeIrSn indicates that the origin of magnetic correlations is different from those in the conventional heavy-fermion metamagnets.
The existence of such an exotic magnetic state deep inside the valence fluctuating regime may be caused by the competition between Kondo singlet formation and geometrical frustration in the ZrNiAl structure.

%


In this paper, we have provided spectroscopic evidence for the strongly intermediate valent character of Ce in the quasikagome Kondo lattice CeIrSn, and reported anomalous negative in-plane linear thermal expansion below 2\,K and negative volume magnetostriction at fields below a metamagnetic crossover that arises at energies two orders of magnitude below the bare Kondo scale.
This behavior is unexpected for an ordinary intermediate valence state and points towards AF correlations, that are polarized beyond $B_{\mathrm{M}}$. The ZF-$\mu $SR relaxation has two components. The relaxation rate of the slow component exhibits a peak around 0.75\,K, while that of the fast component (developing below 2\,K)  saturates at this temperature, which coincides with the broad minimum of the negative $a$-axis thermal expansion.
While $\mu$SR  provides microscopic evidence for a broad distribution of quasi-static internal magnetic fields, thermal expansion and magnetostriction indicate that this is related to antiferromagnetism of bulk nature, incompatible with only a minor volume fraction.
 By comparison with isostructural CeRhSn, we conclude that magnetic frustration, resulting from the quasikagome structure, likely counteracts Kondo singlet formation even in this strongly intermediate valence material.

We would like to thank Stewart Parker and Rob Bewley for an enlightening discussion on the inelastic neutron scattering results. This work was supported by projects JSPS KAKENHI No. JP17K05545, No. JP15H05886, No. JP18KK0078, and No. JP18H01182
 and by the German Research Foundation (DFG) via the Project No. 107745057 (TRR80).
YS thanks the Iketani Science and Technology Foundation (No. 0301080-A and No. 0321131-A) of Japanese funding support.
DTA would like to thank JSPS for invitation fellowship and the Royal Society of London for the International exchange, UK-Japan, funding.
AB would like to acknowledge the Department of Science and Technology (DST) India,
 for an Inspire Faculty Research Grant (DST/INSPIRE/04/2015/000169) and JNCASR for funding support.
AMS thanks the SA-NRF and the URC/FRC of UJ for financial assistance.
We acknowledge the ISIS Facility for beam time at instruments EMU, RB1810852 \cite{ISIS} and MERLIN, RB1990226. We acknowledge DESY (Hamburg, Germany), a member of the Helmholtz Association HGF, for the provision of experimental facilities.


\bibliography{bib_CeIrSn_Dilatometer}

\begin{thebibliography}{47}%
\makeatletter
\providecommand \@ifxundefined [1]{%
 \@ifx{#1\undefined}
}%
\providecommand \@ifnum [1]{%
 \ifnum #1\expandafter \@firstoftwo
 \else \expandafter \@secondoftwo
 \fi
}%
\providecommand \@ifx [1]{%
 \ifx #1\expandafter \@firstoftwo
 \else \expandafter \@secondoftwo
 \fi
}%
\providecommand \natexlab [1]{#1}%
\providecommand \enquote  [1]{``#1''}%
\providecommand \bibnamefont  [1]{#1}%
\providecommand \bibfnamefont [1]{#1}%
\providecommand \citenamefont [1]{#1}%
\providecommand \href@noop [0]{\@secondoftwo}%
\providecommand \href [0]{\begingroup \@sanitize@url \@href}%
\providecommand \@href[1]{\@@startlink{#1}\@@href}%
\providecommand \@@href[1]{\endgroup#1\@@endlink}%
\providecommand \@sanitize@url [0]{\catcode `\\12\catcode `\$12\catcode
  `\&12\catcode `\#12\catcode `\^12\catcode `\_12\catcode `\%12\relax}%
\providecommand \@@startlink[1]{}%
\providecommand \@@endlink[0]{}%
\providecommand \url  [0]{\begingroup\@sanitize@url \@url }%
\providecommand \@url [1]{\endgroup\@href {#1}{\urlprefix }}%
\providecommand \urlprefix  [0]{URL }%
\providecommand \Eprint [0]{\href }%
\providecommand \doibase [0]{http://dx.doi.org/}%
\providecommand \selectlanguage [0]{\@gobble}%
\providecommand \bibinfo  [0]{\@secondoftwo}%
\providecommand \bibfield  [0]{\@secondoftwo}%
\providecommand \translation [1]{[#1]}%
\providecommand \BibitemOpen [0]{}%
\providecommand \bibitemStop [0]{}%
\providecommand \bibitemNoStop [0]{.\EOS\space}%
\providecommand \EOS [0]{\spacefactor3000\relax}%
\providecommand \BibitemShut  [1]{\csname bibitem#1\endcsname}%
\let\auto@bib@innerbib\@empty
\bibitem [{\citenamefont {Loewenhaupt}\ and\ \citenamefont
  {Fischer}()}]{ValenceBook}%
  \BibitemOpen
  \bibfield  {author} {\bibinfo {author} {\bibfnamefont {M.}~\bibnamefont
  {Loewenhaupt}}\ and\ \bibinfo {author} {\bibfnamefont {K.}~\bibnamefont
  {Fischer}},\ }\href@noop {} {}\bibinfo {note} {\textit{{H}andbook on the
  {P}hysics and {C}hemistry of {R}are {E}arths}, {V}ol. 16, {C}hap. 105,
  \textit{{V}alence {F}luctuation and {H}eavy-{F}ermion 4f {S}ystems}
  (1993)}\BibitemShut {NoStop}%
\bibitem [{\citenamefont {Hafner}(1985)}]{Valence_Hafner85}%
  \BibitemOpen
  \bibfield  {author} {\bibinfo {author} {\bibfnamefont {H.}~\bibnamefont
  {Hafner}},\ }\href@noop {} {\bibfield  {journal} {\bibinfo  {journal} {J.
  Magn. Magn. Mater.}\ }\textbf {\bibinfo {volume} {47}},\ \bibinfo {pages}
  {299} (\bibinfo {year} {1985})}\BibitemShut {NoStop}%
\bibitem [{\citenamefont {Edelstein}\ and\ \citenamefont
  {Koon}(1983)}]{CeSn3_Edelstein83}%
  \BibitemOpen
  \bibfield  {author} {\bibinfo {author} {\bibfnamefont {A.}~\bibnamefont
  {Edelstein}}\ and\ \bibinfo {author} {\bibfnamefont {N.}~\bibnamefont
  {Koon}},\ }\href@noop {} {\bibfield  {journal} {\bibinfo  {journal} {Solid
  State Commun.}\ }\textbf {\bibinfo {volume} {48}},\ \bibinfo {pages} {269}
  (\bibinfo {year} {1983})}\BibitemShut {NoStop}%
\bibitem [{\citenamefont {Hulliger}(1993)}]{ZrNiAl_Hulliger93}%
  \BibitemOpen
  \bibfield  {author} {\bibinfo {author} {\bibfnamefont {F.}~\bibnamefont
  {Hulliger}},\ }\href@noop {} {\bibfield  {journal} {\bibinfo  {journal} {J.
  Alloys Compd.}\ }\textbf {\bibinfo {volume} {196}},\ \bibinfo {pages} {225}
  (\bibinfo {year} {1993})}\BibitemShut {NoStop}%
\bibitem [{\citenamefont {Chevalier}\ \emph {et~al.}(2006)\citenamefont
  {Chevalier}, \citenamefont {Sebastian},\ and\ \citenamefont
  {Pottgen}}]{CeTSn_Chevalier06}%
  \BibitemOpen
  \bibfield  {author} {\bibinfo {author} {\bibfnamefont {B.}~\bibnamefont
  {Chevalier}}, \bibinfo {author} {\bibfnamefont {C.}~\bibnamefont
  {Sebastian}}, \ and\ \bibinfo {author} {\bibfnamefont {R.}~\bibnamefont
  {Pottgen}},\ }\href@noop {} {\bibfield  {journal} {\bibinfo  {journal} {Solid
  State Sci.}\ }\textbf {\bibinfo {volume} {8}},\ \bibinfo {pages} {1000}
  (\bibinfo {year} {2006})}\BibitemShut {NoStop}%
\bibitem [{\citenamefont {Pottgen}\ and\ \citenamefont
  {Chevalier}(2015)}]{ZrNiAl_Pottgen15}%
  \BibitemOpen
  \bibfield  {author} {\bibinfo {author} {\bibfnamefont {R.}~\bibnamefont
  {Pottgen}}\ and\ \bibinfo {author} {\bibfnamefont {B.}~\bibnamefont
  {Chevalier}},\ }\href@noop {} {\bibfield  {journal} {\bibinfo  {journal} {Z.
  Naturforsch. B}\ }\textbf {\bibinfo {volume} {70}},\ \bibinfo {pages} {289}
  (\bibinfo {year} {2015})}\BibitemShut {NoStop}%
\bibitem [{\citenamefont {Bando}\ \emph {et~al.}(2000)\citenamefont {Bando},
  \citenamefont {Suemitsu}, \citenamefont {Takagi}, \citenamefont {Tokushima},
  \citenamefont {Echizen}, \citenamefont {Katoh}, \citenamefont {Umeo},
  \citenamefont {Maeda},\ and\ \citenamefont {Takabatake}}]{CeTSn_Bando00}%
  \BibitemOpen
  \bibfield  {author} {\bibinfo {author} {\bibfnamefont {Y.}~\bibnamefont
  {Bando}}, \bibinfo {author} {\bibfnamefont {T.}~\bibnamefont {Suemitsu}},
  \bibinfo {author} {\bibfnamefont {K.}~\bibnamefont {Takagi}}, \bibinfo
  {author} {\bibfnamefont {H.}~\bibnamefont {Tokushima}}, \bibinfo {author}
  {\bibfnamefont {Y.}~\bibnamefont {Echizen}}, \bibinfo {author} {\bibfnamefont
  {K.}~\bibnamefont {Katoh}}, \bibinfo {author} {\bibfnamefont
  {K.}~\bibnamefont {Umeo}}, \bibinfo {author} {\bibfnamefont {Y.}~\bibnamefont
  {Maeda}}, \ and\ \bibinfo {author} {\bibfnamefont {T.}~\bibnamefont
  {Takabatake}},\ }\href@noop {} {\bibfield  {journal} {\bibinfo  {journal} {J.
  Alloys Compd.}\ }\textbf {\bibinfo {volume} {313}},\ \bibinfo {pages} {1 }
  (\bibinfo {year} {2000})}\BibitemShut {NoStop}%
\bibitem [{\citenamefont {Kim}\ \emph {et~al.}(2003)\citenamefont {Kim},
  \citenamefont {Echizen}, \citenamefont {Umeo}, \citenamefont {Kobayashi},
  \citenamefont {Sera}, \citenamefont {Salamakha}, \citenamefont {Sologub},
  \citenamefont {Takabatake}, \citenamefont {Chen}, \citenamefont {Tayama},
  \citenamefont {Sakakibara}, \citenamefont {Jung},\ and\ \citenamefont
  {Maple}}]{CeRhSn_Kim13}%
  \BibitemOpen
  \bibfield  {author} {\bibinfo {author} {\bibfnamefont {M.~S.}\ \bibnamefont
  {Kim}}, \bibinfo {author} {\bibfnamefont {Y.}~\bibnamefont {Echizen}},
  \bibinfo {author} {\bibfnamefont {K.}~\bibnamefont {Umeo}}, \bibinfo {author}
  {\bibfnamefont {S.}~\bibnamefont {Kobayashi}}, \bibinfo {author}
  {\bibfnamefont {M.}~\bibnamefont {Sera}}, \bibinfo {author} {\bibfnamefont
  {P.~S.}\ \bibnamefont {Salamakha}}, \bibinfo {author} {\bibfnamefont {O.~L.}\
  \bibnamefont {Sologub}}, \bibinfo {author} {\bibfnamefont {T.}~\bibnamefont
  {Takabatake}}, \bibinfo {author} {\bibfnamefont {X.}~\bibnamefont {Chen}},
  \bibinfo {author} {\bibfnamefont {T.}~\bibnamefont {Tayama}}, \bibinfo
  {author} {\bibfnamefont {T.}~\bibnamefont {Sakakibara}}, \bibinfo {author}
  {\bibfnamefont {M.~H.}\ \bibnamefont {Jung}}, \ and\ \bibinfo {author}
  {\bibfnamefont {M.~B.}\ \bibnamefont {Maple}},\ }\href@noop {} {\bibfield
  {journal} {\bibinfo  {journal} {Phys. Rev. B}\ }\textbf {\bibinfo {volume}
  {68}},\ \bibinfo {pages} {054416} (\bibinfo {year} {2003})}\BibitemShut
  {NoStop}%
\bibitem [{\citenamefont {Tsuda}\ \emph {et~al.}(2018)\citenamefont {Tsuda},
  \citenamefont {Yang}, \citenamefont {Shimura}, \citenamefont {Umeo},
  \citenamefont {Fukuoka}, \citenamefont {Yamane}, \citenamefont {Onimaru},
  \citenamefont {Takabatake}, \citenamefont {Kikugawa}, \citenamefont
  {Terashima}, \citenamefont {Hirose}, \citenamefont {Uji}, \citenamefont
  {Kittaka},\ and\ \citenamefont {Sakakibara}}]{CeIrSn_Tsuda18}%
  \BibitemOpen
  \bibfield  {author} {\bibinfo {author} {\bibfnamefont {S.}~\bibnamefont
  {Tsuda}}, \bibinfo {author} {\bibfnamefont {C.~L.}\ \bibnamefont {Yang}},
  \bibinfo {author} {\bibfnamefont {Y.}~\bibnamefont {Shimura}}, \bibinfo
  {author} {\bibfnamefont {K.}~\bibnamefont {Umeo}}, \bibinfo {author}
  {\bibfnamefont {H.}~\bibnamefont {Fukuoka}}, \bibinfo {author} {\bibfnamefont
  {Y.}~\bibnamefont {Yamane}}, \bibinfo {author} {\bibfnamefont
  {T.}~\bibnamefont {Onimaru}}, \bibinfo {author} {\bibfnamefont
  {T.}~\bibnamefont {Takabatake}}, \bibinfo {author} {\bibfnamefont
  {N.}~\bibnamefont {Kikugawa}}, \bibinfo {author} {\bibfnamefont
  {T.}~\bibnamefont {Terashima}}, \bibinfo {author} {\bibfnamefont {H.~T.}\
  \bibnamefont {Hirose}}, \bibinfo {author} {\bibfnamefont {S.}~\bibnamefont
  {Uji}}, \bibinfo {author} {\bibfnamefont {S.}~\bibnamefont {Kittaka}}, \ and\
  \bibinfo {author} {\bibfnamefont {T.}~\bibnamefont {Sakakibara}},\
  }\href@noop {} {\bibfield  {journal} {\bibinfo  {journal} {Phys. Rev. B}\
  }\textbf {\bibinfo {volume} {98}},\ \bibinfo {pages} {155147} (\bibinfo
  {year} {2018})}\BibitemShut {NoStop}%
\bibitem [{\citenamefont {Tou}\ \emph {et~al.}(2004)\citenamefont {Tou},
  \citenamefont {Kim}, \citenamefont {Takabatake},\ and\ \citenamefont
  {Sera}}]{CeRhSn_Tou04}%
  \BibitemOpen
  \bibfield  {author} {\bibinfo {author} {\bibfnamefont {H.}~\bibnamefont
  {Tou}}, \bibinfo {author} {\bibfnamefont {M.~S.}\ \bibnamefont {Kim}},
  \bibinfo {author} {\bibfnamefont {T.}~\bibnamefont {Takabatake}}, \ and\
  \bibinfo {author} {\bibfnamefont {M.}~\bibnamefont {Sera}},\ }\href@noop {}
  {\bibfield  {journal} {\bibinfo  {journal} {Phys. Rev. B}\ }\textbf {\bibinfo
  {volume} {70}},\ \bibinfo {pages} {100407(R)} (\bibinfo {year}
  {2004})}\BibitemShut {NoStop}%
\bibitem [{\citenamefont {Schenck}\ \emph {et~al.}(2004)\citenamefont
  {Schenck}, \citenamefont {N.~Gygax}, \citenamefont {S.~Kim},\ and\
  \citenamefont {Takabatake}}]{CeRhSn_Schenck04}%
  \BibitemOpen
  \bibfield  {author} {\bibinfo {author} {\bibfnamefont {A.}~\bibnamefont
  {Schenck}}, \bibinfo {author} {\bibfnamefont {F.}~\bibnamefont {N.~Gygax}},
  \bibinfo {author} {\bibfnamefont {M.}~\bibnamefont {S.~Kim}}, \ and\ \bibinfo
  {author} {\bibfnamefont {T.}~\bibnamefont {Takabatake}},\ }\href@noop {}
  {\bibfield  {journal} {\bibinfo  {journal} {J. Phys. Soc. Jpn}\ }\textbf
  {\bibinfo {volume} {73}},\ \bibinfo {pages} {3099} (\bibinfo {year}
  {2004})}\BibitemShut {NoStop}%
\bibitem [{\citenamefont {Yang}\ \emph {et~al.}(2017)\citenamefont {Yang},
  \citenamefont {Tsuda}, \citenamefont {Umeo}, \citenamefont {Yamane},
  \citenamefont {Onimaru}, \citenamefont {Takabatake}, \citenamefont
  {Kikugawa}, \citenamefont {Terashima},\ and\ \citenamefont
  {Uji}}]{CeRhSn_Yang17}%
  \BibitemOpen
  \bibfield  {author} {\bibinfo {author} {\bibfnamefont {C.~L.}\ \bibnamefont
  {Yang}}, \bibinfo {author} {\bibfnamefont {S.}~\bibnamefont {Tsuda}},
  \bibinfo {author} {\bibfnamefont {K.}~\bibnamefont {Umeo}}, \bibinfo {author}
  {\bibfnamefont {Y.}~\bibnamefont {Yamane}}, \bibinfo {author} {\bibfnamefont
  {T.}~\bibnamefont {Onimaru}}, \bibinfo {author} {\bibfnamefont
  {T.}~\bibnamefont {Takabatake}}, \bibinfo {author} {\bibfnamefont
  {N.}~\bibnamefont {Kikugawa}}, \bibinfo {author} {\bibfnamefont
  {T.}~\bibnamefont {Terashima}}, \ and\ \bibinfo {author} {\bibfnamefont
  {S.}~\bibnamefont {Uji}},\ }\href@noop {} {\bibfield  {journal} {\bibinfo
  {journal} {Phys. Rev. B}\ }\textbf {\bibinfo {volume} {96}},\ \bibinfo
  {pages} {045139} (\bibinfo {year} {2017})}\BibitemShut {NoStop}%
\bibitem [{\citenamefont {Tokiwa}\ \emph {et~al.}(2015)\citenamefont {Tokiwa},
  \citenamefont {Stingl}, \citenamefont {Kim}, \citenamefont {Takabatake},\
  and\ \citenamefont {Gegenwart}}]{CeRhSn_Tokiwa15}%
  \BibitemOpen
  \bibfield  {author} {\bibinfo {author} {\bibfnamefont {Y.}~\bibnamefont
  {Tokiwa}}, \bibinfo {author} {\bibfnamefont {C.}~\bibnamefont {Stingl}},
  \bibinfo {author} {\bibfnamefont {M.-S.}\ \bibnamefont {Kim}}, \bibinfo
  {author} {\bibfnamefont {T.}~\bibnamefont {Takabatake}}, \ and\ \bibinfo
  {author} {\bibfnamefont {P.}~\bibnamefont {Gegenwart}},\ }\href@noop {}
  {\bibfield  {journal} {\bibinfo  {journal} {Sci. Adv.}\ }\textbf {\bibinfo
  {volume} {1}},\ \bibinfo {pages} {1500001} (\bibinfo {year}
  {2015})}\BibitemShut {NoStop}%
\bibitem [{\citenamefont {Takeuchi}\ \emph {et~al.}(2010)\citenamefont
  {Takeuchi}, \citenamefont {Yasui}, \citenamefont {Toda}, \citenamefont
  {Matsushita}, \citenamefont {Yoshiuchi}, \citenamefont {Ohya}, \citenamefont
  {Katayama}, \citenamefont {Hirose}, \citenamefont {Yoshitani}, \citenamefont
  {Honda}, \citenamefont {Sugiyama}, \citenamefont {Hagiwara}, \citenamefont
  {Kindo}, \citenamefont {Yamamoto}, \citenamefont {Haga}, \citenamefont
  {Tanaka}, \citenamefont {Kubo}, \citenamefont {Settai},\ and\ \citenamefont
  {Onuki}}]{YbIr2Zn20_Takeuchi10}%
  \BibitemOpen
  \bibfield  {author} {\bibinfo {author} {\bibfnamefont {T.}~\bibnamefont
  {Takeuchi}}, \bibinfo {author} {\bibfnamefont {S.}~\bibnamefont {Yasui}},
  \bibinfo {author} {\bibfnamefont {M.}~\bibnamefont {Toda}}, \bibinfo {author}
  {\bibfnamefont {M.}~\bibnamefont {Matsushita}}, \bibinfo {author}
  {\bibfnamefont {S.}~\bibnamefont {Yoshiuchi}}, \bibinfo {author}
  {\bibfnamefont {M.}~\bibnamefont {Ohya}}, \bibinfo {author} {\bibfnamefont
  {K.}~\bibnamefont {Katayama}}, \bibinfo {author} {\bibfnamefont
  {Y.}~\bibnamefont {Hirose}}, \bibinfo {author} {\bibfnamefont
  {N.}~\bibnamefont {Yoshitani}}, \bibinfo {author} {\bibfnamefont
  {F.}~\bibnamefont {Honda}}, \bibinfo {author} {\bibfnamefont
  {K.}~\bibnamefont {Sugiyama}}, \bibinfo {author} {\bibfnamefont
  {M.}~\bibnamefont {Hagiwara}}, \bibinfo {author} {\bibfnamefont
  {K.}~\bibnamefont {Kindo}}, \bibinfo {author} {\bibfnamefont
  {E.}~\bibnamefont {Yamamoto}}, \bibinfo {author} {\bibfnamefont
  {Y.}~\bibnamefont {Haga}}, \bibinfo {author} {\bibfnamefont {T.}~\bibnamefont
  {Tanaka}}, \bibinfo {author} {\bibfnamefont {Y.}~\bibnamefont {Kubo}},
  \bibinfo {author} {\bibfnamefont {R.}~\bibnamefont {Settai}}, \ and\ \bibinfo
  {author} {\bibfnamefont {Y.}~\bibnamefont {Onuki}},\ }\href@noop {}
  {\bibfield  {journal} {\bibinfo  {journal} {J. Phys. Soc. Jpn.}\ }\textbf
  {\bibinfo {volume} {79}},\ \bibinfo {pages} {064609} (\bibinfo {year}
  {2010})}\BibitemShut {NoStop}%
\bibitem [{\citenamefont {Garst}\ and\ \citenamefont
  {Rosch}(2005)}]{QCP_Garst05}%
  \BibitemOpen
  \bibfield  {author} {\bibinfo {author} {\bibfnamefont {M.}~\bibnamefont
  {Garst}}\ and\ \bibinfo {author} {\bibfnamefont {A.}~\bibnamefont {Rosch}},\
  }\href@noop {} {\bibfield  {journal} {\bibinfo  {journal} {Phys. Rev. B}\
  }\textbf {\bibinfo {volume} {72}},\ \bibinfo {pages} {205129} (\bibinfo
  {year} {2005})}\BibitemShut {NoStop}%
\bibitem [{\citenamefont {Donath}\ \emph {et~al.}(2009)\citenamefont {Donath},
  \citenamefont {Steglich}, \citenamefont {Bauer}, \citenamefont {Ronning},
  \citenamefont {Sarrao},\ and\ \citenamefont {Gegenwart}}]{CeRhIn5_Donath09}%
  \BibitemOpen
  \bibfield  {author} {\bibinfo {author} {\bibfnamefont {J.~G.}\ \bibnamefont
  {Donath}}, \bibinfo {author} {\bibfnamefont {F.}~\bibnamefont {Steglich}},
  \bibinfo {author} {\bibfnamefont {E.~D.}\ \bibnamefont {Bauer}}, \bibinfo
  {author} {\bibfnamefont {F.}~\bibnamefont {Ronning}}, \bibinfo {author}
  {\bibfnamefont {J.~L.}\ \bibnamefont {Sarrao}}, \ and\ \bibinfo {author}
  {\bibfnamefont {P.}~\bibnamefont {Gegenwart}},\ }\href@noop {} {\bibfield
  {journal} {\bibinfo  {journal} {Europhys. Lett.}\ }\textbf {\bibinfo {volume}
  {87}},\ \bibinfo {pages} {57011} (\bibinfo {year} {2009})}\BibitemShut
  {NoStop}%
\bibitem [{\citenamefont {Grube}\ \emph {et~al.}(2018)\citenamefont {Grube},
  \citenamefont {Pintschovius}, \citenamefont {Weber}, \citenamefont
  {Castellan}, \citenamefont {Zaum}, \citenamefont {Kuntz}, \citenamefont
  {Schweiss}, \citenamefont {Stockert}, \citenamefont {Bachus}, \citenamefont
  {Shimura}, \citenamefont {Fritsch},\ and\ \citenamefont
  {L\"ohneysen}}]{CeCu6_Grube18}%
  \BibitemOpen
  \bibfield  {author} {\bibinfo {author} {\bibfnamefont {K.}~\bibnamefont
  {Grube}}, \bibinfo {author} {\bibfnamefont {L.}~\bibnamefont {Pintschovius}},
  \bibinfo {author} {\bibfnamefont {F.}~\bibnamefont {Weber}}, \bibinfo
  {author} {\bibfnamefont {J.-P.}\ \bibnamefont {Castellan}}, \bibinfo {author}
  {\bibfnamefont {S.}~\bibnamefont {Zaum}}, \bibinfo {author} {\bibfnamefont
  {S.}~\bibnamefont {Kuntz}}, \bibinfo {author} {\bibfnamefont
  {P.}~\bibnamefont {Schweiss}}, \bibinfo {author} {\bibfnamefont
  {O.}~\bibnamefont {Stockert}}, \bibinfo {author} {\bibfnamefont
  {S.}~\bibnamefont {Bachus}}, \bibinfo {author} {\bibfnamefont
  {Y.}~\bibnamefont {Shimura}}, \bibinfo {author} {\bibfnamefont
  {V.}~\bibnamefont {Fritsch}}, \ and\ \bibinfo {author} {\bibfnamefont
  {H.~v.}\ \bibnamefont {L\"ohneysen}},\ }\href@noop {} {\bibfield  {journal}
  {\bibinfo  {journal} {Phys. Rev. Lett.}\ }\textbf {\bibinfo {volume} {121}},\
  \bibinfo {pages} {087203} (\bibinfo {year} {2018})}\BibitemShut {NoStop}%
\bibitem [{\citenamefont {K\"uchler}\ \emph {et~al.}(2006)\citenamefont
  {K\"uchler}, \citenamefont {Gegenwart}, \citenamefont {Custers},
  \citenamefont {Stockert}, \citenamefont {Caroca-Canales}, \citenamefont
  {Geibel}, \citenamefont {Sereni},\ and\ \citenamefont
  {Steglich}}]{CeSn3_Kuchler06}%
  \BibitemOpen
  \bibfield  {author} {\bibinfo {author} {\bibfnamefont {R.}~\bibnamefont
  {K\"uchler}}, \bibinfo {author} {\bibfnamefont {P.}~\bibnamefont
  {Gegenwart}}, \bibinfo {author} {\bibfnamefont {J.}~\bibnamefont {Custers}},
  \bibinfo {author} {\bibfnamefont {O.}~\bibnamefont {Stockert}}, \bibinfo
  {author} {\bibfnamefont {N.}~\bibnamefont {Caroca-Canales}}, \bibinfo
  {author} {\bibfnamefont {C.}~\bibnamefont {Geibel}}, \bibinfo {author}
  {\bibfnamefont {J.~G.}\ \bibnamefont {Sereni}}, \ and\ \bibinfo {author}
  {\bibfnamefont {F.}~\bibnamefont {Steglich}},\ }\href@noop {} {\bibfield
  {journal} {\bibinfo  {journal} {Phys. Rev. Lett.}\ }\textbf {\bibinfo
  {volume} {96}},\ \bibinfo {pages} {256403} (\bibinfo {year}
  {2006})}\BibitemShut {NoStop}%
\bibitem [{\citenamefont {K\"uchler}\ \emph
  {et~al.}(2017{\natexlab{a}})\citenamefont {K\"uchler}, \citenamefont
  {Stingl}, \citenamefont {Tokiwa}, \citenamefont {Kim}, \citenamefont
  {Takabatake},\ and\ \citenamefont {Gegenwart}}]{CeRhSn_Kuchler17}%
  \BibitemOpen
  \bibfield  {author} {\bibinfo {author} {\bibfnamefont {R.}~\bibnamefont
  {K\"uchler}}, \bibinfo {author} {\bibfnamefont {C.}~\bibnamefont {Stingl}},
  \bibinfo {author} {\bibfnamefont {Y.}~\bibnamefont {Tokiwa}}, \bibinfo
  {author} {\bibfnamefont {M.~S.}\ \bibnamefont {Kim}}, \bibinfo {author}
  {\bibfnamefont {T.}~\bibnamefont {Takabatake}}, \ and\ \bibinfo {author}
  {\bibfnamefont {P.}~\bibnamefont {Gegenwart}},\ }\href@noop {} {\bibfield
  {journal} {\bibinfo  {journal} {Phys. Rev. B}\ }\textbf {\bibinfo {volume}
  {96}},\ \bibinfo {pages} {241110(R)} (\bibinfo {year}
  {2017}{\natexlab{a}})}\BibitemShut {NoStop}%
\bibitem [{\citenamefont {Oyamada}\ \emph {et~al.}(2008)\citenamefont
  {Oyamada}, \citenamefont {Maegawa}, \citenamefont {Nishiyama}, \citenamefont
  {Kitazawa},\ and\ \citenamefont {Isikawa}}]{CePdAl_Oyamada08}%
  \BibitemOpen
  \bibfield  {author} {\bibinfo {author} {\bibfnamefont {A.}~\bibnamefont
  {Oyamada}}, \bibinfo {author} {\bibfnamefont {S.}~\bibnamefont {Maegawa}},
  \bibinfo {author} {\bibfnamefont {M.}~\bibnamefont {Nishiyama}}, \bibinfo
  {author} {\bibfnamefont {H.}~\bibnamefont {Kitazawa}}, \ and\ \bibinfo
  {author} {\bibfnamefont {Y.}~\bibnamefont {Isikawa}},\ }\href@noop {}
  {\bibfield  {journal} {\bibinfo  {journal} {Phys. Rev. B}\ }\textbf {\bibinfo
  {volume} {77}},\ \bibinfo {pages} {064432} (\bibinfo {year}
  {2008})}\BibitemShut {NoStop}%
\bibitem [{\citenamefont {Donni}\ \emph {et~al.}(1996)\citenamefont {Donni},
  \citenamefont {Ehlers}, \citenamefont {Maletta}, \citenamefont {Fischer},
  \citenamefont {Kitazawa},\ and\ \citenamefont {Zolliker}}]{CePdAl_Donni96}%
  \BibitemOpen
  \bibfield  {author} {\bibinfo {author} {\bibfnamefont {A.}~\bibnamefont
  {Donni}}, \bibinfo {author} {\bibfnamefont {G.}~\bibnamefont {Ehlers}},
  \bibinfo {author} {\bibfnamefont {H.}~\bibnamefont {Maletta}}, \bibinfo
  {author} {\bibfnamefont {P.}~\bibnamefont {Fischer}}, \bibinfo {author}
  {\bibfnamefont {H.}~\bibnamefont {Kitazawa}}, \ and\ \bibinfo {author}
  {\bibfnamefont {M.}~\bibnamefont {Zolliker}},\ }\href@noop {} {\bibfield
  {journal} {\bibinfo  {journal} {J. Phys.: Condens. Matter}\ }\textbf
  {\bibinfo {volume} {8}},\ \bibinfo {pages} {11213} (\bibinfo {year}
  {1996})}\BibitemShut {NoStop}%
\bibitem [{\citenamefont {Fang}\ \emph {et~al.}(2019)\citenamefont {Fang},
  \citenamefont {Dunsiger}, \citenamefont {Pal}, \citenamefont {Akintola},
  \citenamefont {Sonier},\ and\ \citenamefont {Mun}}]{CePtPb_Fang19}%
  \BibitemOpen
  \bibfield  {author} {\bibinfo {author} {\bibfnamefont {A.~C.~Y.}\
  \bibnamefont {Fang}}, \bibinfo {author} {\bibfnamefont {S.~R.}\ \bibnamefont
  {Dunsiger}}, \bibinfo {author} {\bibfnamefont {A.}~\bibnamefont {Pal}},
  \bibinfo {author} {\bibfnamefont {K.}~\bibnamefont {Akintola}}, \bibinfo
  {author} {\bibfnamefont {J.~E.}\ \bibnamefont {Sonier}}, \ and\ \bibinfo
  {author} {\bibfnamefont {E.}~\bibnamefont {Mun}},\ }\href@noop {} {\bibfield
  {journal} {\bibinfo  {journal} {Phys. Rev. B}\ }\textbf {\bibinfo {volume}
  {100}},\ \bibinfo {pages} {024404} (\bibinfo {year} {2019})}\BibitemShut
  {NoStop}%
\bibitem [{SM()}]{SM}%
  \BibitemOpen
  \href@noop {} {}\bibinfo {note} {See {S}upplemental {M}aterial, which
  includes {R}efs. \cite{Pratt00, CeRhGe3, CePd3_Murani96, Jones91,
  Stephenson88, Maisuradze18} at [{URL} will be inserted by publisher] for
  experimental details, {HAXPES}, {INS}, {M}agnetostriction and
  {TF}-$\mu${SR}}\BibitemShut {NoStop}%
\bibitem [{\citenamefont {Pratt}(2000)}]{Pratt00}%
  \BibitemOpen
  \bibfield  {author} {\bibinfo {author} {\bibfnamefont {F.}~\bibnamefont
  {Pratt}},\ }\href@noop {} {\bibfield  {journal} {\bibinfo  {journal} {Physica
  B: Condens. Matter}\ }\textbf {\bibinfo {volume} {289}},\ \bibinfo {pages}
  {710} (\bibinfo {year} {2000})}\BibitemShut {NoStop}%
\bibitem [{\citenamefont {Hillier}\ \emph {et~al.}(2012)\citenamefont
  {Hillier}, \citenamefont {Adroja}, \citenamefont {Manuel}, \citenamefont
  {Anand}, \citenamefont {Taylor}, \citenamefont {McEwen}, \citenamefont
  {Rainford},\ and\ \citenamefont {Koza}}]{CeRhGe3}%
  \BibitemOpen
  \bibfield  {author} {\bibinfo {author} {\bibfnamefont {A.~D.}\ \bibnamefont
  {Hillier}}, \bibinfo {author} {\bibfnamefont {D.~T.}\ \bibnamefont {Adroja}},
  \bibinfo {author} {\bibfnamefont {P.}~\bibnamefont {Manuel}}, \bibinfo
  {author} {\bibfnamefont {V.~K.}\ \bibnamefont {Anand}}, \bibinfo {author}
  {\bibfnamefont {J.~W.}\ \bibnamefont {Taylor}}, \bibinfo {author}
  {\bibfnamefont {K.~A.}\ \bibnamefont {McEwen}}, \bibinfo {author}
  {\bibfnamefont {B.~D.}\ \bibnamefont {Rainford}}, \ and\ \bibinfo {author}
  {\bibfnamefont {M.~M.}\ \bibnamefont {Koza}},\ }\href {\doibase
  10.1103/PhysRevB.85.134405} {\bibfield  {journal} {\bibinfo  {journal} {Phys.
  Rev. B}\ }\textbf {\bibinfo {volume} {85}},\ \bibinfo {pages} {134405}
  (\bibinfo {year} {2012})}\BibitemShut {NoStop}%
\bibitem [{\citenamefont {Murani}\ \emph {et~al.}(1996)\citenamefont {Murani},
  \citenamefont {Severing},\ and\ \citenamefont {Marshall}}]{CePd3_Murani96}%
  \BibitemOpen
  \bibfield  {author} {\bibinfo {author} {\bibfnamefont {A.~P.}\ \bibnamefont
  {Murani}}, \bibinfo {author} {\bibfnamefont {A.}~\bibnamefont {Severing}}, \
  and\ \bibinfo {author} {\bibfnamefont {W.~G.}\ \bibnamefont {Marshall}},\
  }\href@noop {} {\bibfield  {journal} {\bibinfo  {journal} {Phys. Rev. B}\
  }\textbf {\bibinfo {volume} {53}},\ \bibinfo {pages} {2641} (\bibinfo {year}
  {1996})}\BibitemShut {NoStop}%
\bibitem [{\citenamefont {Jones}\ and\ \citenamefont {Hore}(1991)}]{Jones91}%
  \BibitemOpen
  \bibfield  {author} {\bibinfo {author} {\bibfnamefont {J.}~\bibnamefont
  {Jones}}\ and\ \bibinfo {author} {\bibfnamefont {P.}~\bibnamefont {Hore}},\
  }\href@noop {} {\bibfield  {journal} {\bibinfo  {journal} {J. Magn. Res.}\
  }\textbf {\bibinfo {volume} {92}},\ \bibinfo {pages} {276} (\bibinfo {year}
  {1991})}\BibitemShut {NoStop}%
\bibitem [{\citenamefont {Stephenson}(1988)}]{Stephenson88}%
  \BibitemOpen
  \bibfield  {author} {\bibinfo {author} {\bibfnamefont {D.~S.}\ \bibnamefont
  {Stephenson}},\ }\href@noop {} {\bibfield  {journal} {\bibinfo  {journal}
  {Progress in NMR Spectroscopy}\ }\textbf {\bibinfo {volume} {20}},\ \bibinfo
  {pages} {515} (\bibinfo {year} {1988})}\BibitemShut {NoStop}%
\bibitem [{\citenamefont {Maisuradze}\ \emph {et~al.}(2018)\citenamefont
  {Maisuradze}, \citenamefont {Yaouanc},\ and\ \citenamefont
  {de~Reotier}}]{Maisuradze18}%
  \BibitemOpen
  \bibfield  {author} {\bibinfo {author} {\bibfnamefont {A.}~\bibnamefont
  {Maisuradze}}, \bibinfo {author} {\bibfnamefont {A.}~\bibnamefont {Yaouanc}},
  \ and\ \bibinfo {author} {\bibfnamefont {P.~D.}\ \bibnamefont {de~Reotier}},\
  }\href@noop {} {\bibfield  {journal} {\bibinfo  {journal} {JPS Conf. Proc.}\
  }\textbf {\bibinfo {volume} {21}},\ \bibinfo {pages} {011053} (\bibinfo
  {year} {2018})}\BibitemShut {NoStop}%
\bibitem [{\citenamefont {Strigari}\ \emph {et~al.}(2015)\citenamefont
  {Strigari}, \citenamefont {Sundermann}, \citenamefont {Muro}, \citenamefont
  {Yutani}, \citenamefont {Takabatake}, \citenamefont {Tsuei}, \citenamefont
  {Liao}, \citenamefont {Tanaka}, \citenamefont {Thalmeier}, \citenamefont
  {Haverkort}, \citenamefont {Tjeng},\ and\ \citenamefont
  {Severing}}]{Strigari2015}%
  \BibitemOpen
  \bibfield  {author} {\bibinfo {author} {\bibfnamefont {F.}~\bibnamefont
  {Strigari}}, \bibinfo {author} {\bibfnamefont {M.}~\bibnamefont
  {Sundermann}}, \bibinfo {author} {\bibfnamefont {Y.}~\bibnamefont {Muro}},
  \bibinfo {author} {\bibfnamefont {K.}~\bibnamefont {Yutani}}, \bibinfo
  {author} {\bibfnamefont {T.}~\bibnamefont {Takabatake}}, \bibinfo {author}
  {\bibfnamefont {K.-D.}\ \bibnamefont {Tsuei}}, \bibinfo {author}
  {\bibfnamefont {Y.}~\bibnamefont {Liao}}, \bibinfo {author} {\bibfnamefont
  {A.}~\bibnamefont {Tanaka}}, \bibinfo {author} {\bibfnamefont
  {P.}~\bibnamefont {Thalmeier}}, \bibinfo {author} {\bibfnamefont
  {M.}~\bibnamefont {Haverkort}}, \bibinfo {author} {\bibfnamefont
  {L.}~\bibnamefont {Tjeng}}, \ and\ \bibinfo {author} {\bibfnamefont
  {A.}~\bibnamefont {Severing}},\ }\href {\doibase
  https://doi.org/10.1016/j.elspec.2015.01.004} {\bibfield  {journal} {\bibinfo
   {journal} {J. Elec. Spect. and Rel. Phen.}\ }\textbf {\bibinfo {volume}
  {199}},\ \bibinfo {pages} {56} (\bibinfo {year} {2015})}\BibitemShut
  {NoStop}%
\bibitem [{\citenamefont {Sundermann}\ \emph {et~al.}(2016)\citenamefont
  {Sundermann}, \citenamefont {Strigari}, \citenamefont {Willers},
  \citenamefont {Weinen}, \citenamefont {Liao}, \citenamefont {Tsuei},
  \citenamefont {Hiraoka}, \citenamefont {Ishii}, \citenamefont {Yamaoka},
  \citenamefont {Mizuki}, \citenamefont {Zekko}, \citenamefont {Bauer},
  \citenamefont {Sarrao}, \citenamefont {Thompson}, \citenamefont {Lejay},
  \citenamefont {Muro}, \citenamefont {Yutani}, \citenamefont {Takabatake},
  \citenamefont {Tanaka}, \citenamefont {Hollmann}, \citenamefont {Tjeng},\
  and\ \citenamefont {Severing}}]{Sundermann2016}%
  \BibitemOpen
  \bibfield  {author} {\bibinfo {author} {\bibfnamefont {M.}~\bibnamefont
  {Sundermann}}, \bibinfo {author} {\bibfnamefont {F.}~\bibnamefont
  {Strigari}}, \bibinfo {author} {\bibfnamefont {T.}~\bibnamefont {Willers}},
  \bibinfo {author} {\bibfnamefont {J.}~\bibnamefont {Weinen}}, \bibinfo
  {author} {\bibfnamefont {Y.}~\bibnamefont {Liao}}, \bibinfo {author}
  {\bibfnamefont {K.-D.}\ \bibnamefont {Tsuei}}, \bibinfo {author}
  {\bibfnamefont {N.}~\bibnamefont {Hiraoka}}, \bibinfo {author} {\bibfnamefont
  {H.}~\bibnamefont {Ishii}}, \bibinfo {author} {\bibfnamefont
  {H.}~\bibnamefont {Yamaoka}}, \bibinfo {author} {\bibfnamefont
  {J.}~\bibnamefont {Mizuki}}, \bibinfo {author} {\bibfnamefont
  {Y.}~\bibnamefont {Zekko}}, \bibinfo {author} {\bibfnamefont
  {E.}~\bibnamefont {Bauer}}, \bibinfo {author} {\bibfnamefont
  {J.}~\bibnamefont {Sarrao}}, \bibinfo {author} {\bibfnamefont
  {J.}~\bibnamefont {Thompson}}, \bibinfo {author} {\bibfnamefont
  {P.}~\bibnamefont {Lejay}}, \bibinfo {author} {\bibfnamefont
  {Y.}~\bibnamefont {Muro}}, \bibinfo {author} {\bibfnamefont {K.}~\bibnamefont
  {Yutani}}, \bibinfo {author} {\bibfnamefont {T.}~\bibnamefont {Takabatake}},
  \bibinfo {author} {\bibfnamefont {A.}~\bibnamefont {Tanaka}}, \bibinfo
  {author} {\bibfnamefont {N.}~\bibnamefont {Hollmann}}, \bibinfo {author}
  {\bibfnamefont {L.}~\bibnamefont {Tjeng}}, \ and\ \bibinfo {author}
  {\bibfnamefont {A.}~\bibnamefont {Severing}},\ }\href {\doibase
  https://doi.org/10.1016/j.elspec.2016.02.002} {\bibfield  {journal} {\bibinfo
   {journal} {J. Elec. Spect. and Rel. Phen.}\ }\textbf {\bibinfo {volume}
  {209}},\ \bibinfo {pages} {1} (\bibinfo {year} {2016})}\BibitemShut {NoStop}%
\bibitem [{\citenamefont {Sundermann}\ \emph {et~al.}(2017)\citenamefont
  {Sundermann}, \citenamefont {Chen}, \citenamefont {Utsumi}, \citenamefont
  {Wu}, \citenamefont {Tsuei}, \citenamefont {Haenel}, \citenamefont
  {Prokofiev}, \citenamefont {Paschen}, \citenamefont {Tanaka}, \citenamefont
  {Tjeng},\ and\ \citenamefont {Severing}}]{Sundermann2017}%
  \BibitemOpen
  \bibfield  {author} {\bibinfo {author} {\bibfnamefont {M.}~\bibnamefont
  {Sundermann}}, \bibinfo {author} {\bibfnamefont {K.}~\bibnamefont {Chen}},
  \bibinfo {author} {\bibfnamefont {Y.}~\bibnamefont {Utsumi}}, \bibinfo
  {author} {\bibfnamefont {Y.-H.}\ \bibnamefont {Wu}}, \bibinfo {author}
  {\bibfnamefont {K.-D.}\ \bibnamefont {Tsuei}}, \bibinfo {author}
  {\bibfnamefont {J.}~\bibnamefont {Haenel}}, \bibinfo {author} {\bibfnamefont
  {A.}~\bibnamefont {Prokofiev}}, \bibinfo {author} {\bibfnamefont
  {S.}~\bibnamefont {Paschen}}, \bibinfo {author} {\bibfnamefont
  {A.}~\bibnamefont {Tanaka}}, \bibinfo {author} {\bibfnamefont {L.~H.}\
  \bibnamefont {Tjeng}}, \ and\ \bibinfo {author} {\bibfnamefont
  {A.}~\bibnamefont {Severing}},\ }\href {\doibase
  10.1088/1742-6596/807/2/022001} {\bibfield  {journal} {\bibinfo  {journal}
  {J. Phys.: Conf. Ser.}\ }\textbf {\bibinfo {volume} {807}},\ \bibinfo {pages}
  {022001} (\bibinfo {year} {2017})}\BibitemShut {NoStop}%
\bibitem [{\citenamefont {Imer}\ and\ \citenamefont
  {Wuilloud}(1987)}]{Imer1987}%
  \BibitemOpen
  \bibfield  {author} {\bibinfo {author} {\bibfnamefont {J.~M.}\ \bibnamefont
  {Imer}}\ and\ \bibinfo {author} {\bibfnamefont {E.}~\bibnamefont
  {Wuilloud}},\ }\href {\doibase 10.1007/BF01311650} {\bibfield  {journal}
  {\bibinfo  {journal} {Z. Phys. B. Condens. matter}\ }\textbf {\bibinfo
  {volume} {66}},\ \bibinfo {pages} {153} (\bibinfo {year} {1987})}\BibitemShut
  {NoStop}%
\bibitem [{\citenamefont {Tanaka}\ and\ \citenamefont {Jo}(1994)}]{Tanaka1994}%
  \BibitemOpen
  \bibfield  {author} {\bibinfo {author} {\bibfnamefont {A.}~\bibnamefont
  {Tanaka}}\ and\ \bibinfo {author} {\bibfnamefont {T.}~\bibnamefont {Jo}},\
  }\href@noop {} {\bibfield  {journal} {\bibinfo  {journal} {J. Phys. Soc.
  Jpn.}\ }\textbf {\bibinfo {volume} {63}},\ \bibinfo {pages} {2788} (\bibinfo
  {year} {1994})}\BibitemShut {NoStop}%
\bibitem [{\citenamefont {K\"uchler}\ \emph
  {et~al.}(2017{\natexlab{b}})\citenamefont {K\"uchler}, \citenamefont
  {W\"orl}, \citenamefont {Gegenwart}, \citenamefont {Berben}, \citenamefont
  {Bryant},\ and\ \citenamefont {Wiedmann}}]{Dilatometer_Kuchler17}%
  \BibitemOpen
  \bibfield  {author} {\bibinfo {author} {\bibfnamefont {R.}~\bibnamefont
  {K\"uchler}}, \bibinfo {author} {\bibfnamefont {A.}~\bibnamefont {W\"orl}},
  \bibinfo {author} {\bibfnamefont {P.}~\bibnamefont {Gegenwart}}, \bibinfo
  {author} {\bibfnamefont {M.}~\bibnamefont {Berben}}, \bibinfo {author}
  {\bibfnamefont {B.}~\bibnamefont {Bryant}}, \ and\ \bibinfo {author}
  {\bibfnamefont {S.}~\bibnamefont {Wiedmann}},\ }\href@noop {} {\bibfield
  {journal} {\bibinfo  {journal} {Rev. Sci. Instrum.}\ }\textbf {\bibinfo
  {volume} {88}},\ \bibinfo {pages} {083903} (\bibinfo {year}
  {2017}{\natexlab{b}})}\BibitemShut {NoStop}%
\bibitem [{\citenamefont {Haen}\ \emph {et~al.}(1987)\citenamefont {Haen},
  \citenamefont {Flouquet}, \citenamefont {Lapierre}, \citenamefont {Lejay},\
  and\ \citenamefont {Remenyi}}]{CeRu2Si2_Haen87}%
  \BibitemOpen
  \bibfield  {author} {\bibinfo {author} {\bibfnamefont {P.}~\bibnamefont
  {Haen}}, \bibinfo {author} {\bibfnamefont {J.}~\bibnamefont {Flouquet}},
  \bibinfo {author} {\bibfnamefont {F.}~\bibnamefont {Lapierre}}, \bibinfo
  {author} {\bibfnamefont {P.}~\bibnamefont {Lejay}}, \ and\ \bibinfo {author}
  {\bibfnamefont {G.}~\bibnamefont {Remenyi}},\ }\href@noop {} {\bibfield
  {journal} {\bibinfo  {journal} {J. Low Temp. Phys.}\ }\textbf {\bibinfo
  {volume} {67}},\ \bibinfo {pages} {391} (\bibinfo {year} {1987})}\BibitemShut
  {NoStop}%
\bibitem [{\citenamefont {Lacerda}\ \emph {et~al.}(1989)\citenamefont
  {Lacerda}, \citenamefont {de~Visser}, \citenamefont {Haen}, \citenamefont
  {Lejay},\ and\ \citenamefont {Flouquet}}]{CeRu2Si2_Lacerda89}%
  \BibitemOpen
  \bibfield  {author} {\bibinfo {author} {\bibfnamefont {A.}~\bibnamefont
  {Lacerda}}, \bibinfo {author} {\bibfnamefont {A.}~\bibnamefont {de~Visser}},
  \bibinfo {author} {\bibfnamefont {P.}~\bibnamefont {Haen}}, \bibinfo {author}
  {\bibfnamefont {P.}~\bibnamefont {Lejay}}, \ and\ \bibinfo {author}
  {\bibfnamefont {J.}~\bibnamefont {Flouquet}},\ }\href@noop {} {\bibfield
  {journal} {\bibinfo  {journal} {Phys. Rev. B}\ }\textbf {\bibinfo {volume}
  {40}},\ \bibinfo {pages} {8759} (\bibinfo {year} {1989})}\BibitemShut
  {NoStop}%
\bibitem [{\citenamefont {de~Visser}\ \emph {et~al.}(1991)\citenamefont
  {de~Visser}, \citenamefont {Flouquet}, \citenamefont {Franse}, \citenamefont
  {Haen}, \citenamefont {Hasselbach}, \citenamefont {Lacerda},\ and\
  \citenamefont {Taillefer}}]{CeRu2Si2_Visser91}%
  \BibitemOpen
  \bibfield  {author} {\bibinfo {author} {\bibfnamefont {A.}~\bibnamefont
  {de~Visser}}, \bibinfo {author} {\bibfnamefont {J.}~\bibnamefont {Flouquet}},
  \bibinfo {author} {\bibfnamefont {J.}~\bibnamefont {Franse}}, \bibinfo
  {author} {\bibfnamefont {P.}~\bibnamefont {Haen}}, \bibinfo {author}
  {\bibfnamefont {K.}~\bibnamefont {Hasselbach}}, \bibinfo {author}
  {\bibfnamefont {A.}~\bibnamefont {Lacerda}}, \ and\ \bibinfo {author}
  {\bibfnamefont {L.}~\bibnamefont {Taillefer}},\ }\href@noop {} {\bibfield
  {journal} {\bibinfo  {journal} {Physica B: Condens. Matter}\ }\textbf
  {\bibinfo {volume} {171}},\ \bibinfo {pages} {190} (\bibinfo {year}
  {1991})}\BibitemShut {NoStop}%
\bibitem [{\citenamefont {Puech}\ \emph {et~al.}(1988)\citenamefont {Puech},
  \citenamefont {Mignot}, \citenamefont {Lejay}, \citenamefont {Haen},\ and\
  \citenamefont {Flouquet}}]{CeRu2Si2_Puech88}%
  \BibitemOpen
  \bibfield  {author} {\bibinfo {author} {\bibfnamefont {L.}~\bibnamefont
  {Puech}}, \bibinfo {author} {\bibfnamefont {J.-M.}\ \bibnamefont {Mignot}},
  \bibinfo {author} {\bibfnamefont {P.}~\bibnamefont {Lejay}}, \bibinfo
  {author} {\bibfnamefont {P.}~\bibnamefont {Haen}}, \ and\ \bibinfo {author}
  {\bibfnamefont {J.}~\bibnamefont {Flouquet}},\ }\href@noop {} {\bibfield
  {journal} {\bibinfo  {journal} {J. Low Temp. Phys.}\ }\textbf {\bibinfo
  {volume} {70}},\ \bibinfo {pages} {237} (\bibinfo {year} {1988})}\BibitemShut
  {NoStop}%
\bibitem [{\citenamefont {Matsuhira}\ \emph {et~al.}(1999)\citenamefont
  {Matsuhira}, \citenamefont {Sakakibara}, \citenamefont {Nomachi},
  \citenamefont {Tayama}, \citenamefont {Tenya}, \citenamefont {Amitsuka},
  \citenamefont {Maezawa},\ and\ \citenamefont {Onuki}}]{CeRu2Si2_Matsuhira99}%
  \BibitemOpen
  \bibfield  {author} {\bibinfo {author} {\bibfnamefont {K.}~\bibnamefont
  {Matsuhira}}, \bibinfo {author} {\bibfnamefont {T.}~\bibnamefont
  {Sakakibara}}, \bibinfo {author} {\bibfnamefont {A.}~\bibnamefont {Nomachi}},
  \bibinfo {author} {\bibfnamefont {T.}~\bibnamefont {Tayama}}, \bibinfo
  {author} {\bibfnamefont {K.}~\bibnamefont {Tenya}}, \bibinfo {author}
  {\bibfnamefont {H.}~\bibnamefont {Amitsuka}}, \bibinfo {author}
  {\bibfnamefont {K.}~\bibnamefont {Maezawa}}, \ and\ \bibinfo {author}
  {\bibfnamefont {Y.}~\bibnamefont {Onuki}},\ }\href@noop {} {\bibfield
  {journal} {\bibinfo  {journal} {J. Phys. Soc. Jpn.}\ }\textbf {\bibinfo
  {volume} {68}},\ \bibinfo {pages} {3402} (\bibinfo {year}
  {1999})}\BibitemShut {NoStop}%
\bibitem [{\citenamefont {Kaiser}\ and\ \citenamefont
  {Fulde}(1988)}]{Gruneisen_Kaiser88}%
  \BibitemOpen
  \bibfield  {author} {\bibinfo {author} {\bibfnamefont {A.~B.}\ \bibnamefont
  {Kaiser}}\ and\ \bibinfo {author} {\bibfnamefont {P.}~\bibnamefont {Fulde}},\
  }\href@noop {} {\bibfield  {journal} {\bibinfo  {journal} {Phys. Rev. B}\
  }\textbf {\bibinfo {volume} {37}},\ \bibinfo {pages} {5357} (\bibinfo {year}
  {1988})}\BibitemShut {NoStop}%
\bibitem [{\citenamefont {Tsujii}\ \emph {et~al.}(2001)\citenamefont {Tsujii},
  \citenamefont {Mitamura}, \citenamefont {Goto}, \citenamefont {Yoshimura},
  \citenamefont {Kosuge}, \citenamefont {Terashima}, \citenamefont {Takamasu},
  \citenamefont {Kitazawa}, \citenamefont {Kato},\ and\ \citenamefont
  {Kido}}]{YbCu5_Tsujii01}%
  \BibitemOpen
  \bibfield  {author} {\bibinfo {author} {\bibfnamefont {N.}~\bibnamefont
  {Tsujii}}, \bibinfo {author} {\bibfnamefont {H.}~\bibnamefont {Mitamura}},
  \bibinfo {author} {\bibfnamefont {T.}~\bibnamefont {Goto}}, \bibinfo {author}
  {\bibfnamefont {K.}~\bibnamefont {Yoshimura}}, \bibinfo {author}
  {\bibfnamefont {K.}~\bibnamefont {Kosuge}}, \bibinfo {author} {\bibfnamefont
  {T.}~\bibnamefont {Terashima}}, \bibinfo {author} {\bibfnamefont
  {T.}~\bibnamefont {Takamasu}}, \bibinfo {author} {\bibfnamefont
  {H.}~\bibnamefont {Kitazawa}}, \bibinfo {author} {\bibfnamefont
  {S.}~\bibnamefont {Kato}}, \ and\ \bibinfo {author} {\bibfnamefont
  {G.}~\bibnamefont {Kido}},\ }\href@noop {} {\bibfield  {journal} {\bibinfo
  {journal} {Physica B: Condens. Matter}\ }\textbf {\bibinfo {volume} {294}},\
  \bibinfo {pages} {284} (\bibinfo {year} {2001})}\BibitemShut {NoStop}%
\bibitem [{\citenamefont {Shimura}\ \emph {et~al.}(2011)\citenamefont
  {Shimura}, \citenamefont {Sakakibara}, \citenamefont {Yoshiuchi},
  \citenamefont {Honda}, \citenamefont {Settai},\ and\ \citenamefont
  {Onuki}}]{YbIr2Zn20_Shimura10}%
  \BibitemOpen
  \bibfield  {author} {\bibinfo {author} {\bibfnamefont {Y.}~\bibnamefont
  {Shimura}}, \bibinfo {author} {\bibfnamefont {T.}~\bibnamefont {Sakakibara}},
  \bibinfo {author} {\bibfnamefont {S.}~\bibnamefont {Yoshiuchi}}, \bibinfo
  {author} {\bibfnamefont {F.}~\bibnamefont {Honda}}, \bibinfo {author}
  {\bibfnamefont {R.}~\bibnamefont {Settai}}, \ and\ \bibinfo {author}
  {\bibfnamefont {Y.}~\bibnamefont {Onuki}},\ }\href@noop {} {\bibfield
  {journal} {\bibinfo  {journal} {J. Phys. Soc. Jpn.}\ }\textbf {\bibinfo
  {volume} {80}},\ \bibinfo {pages} {SA051} (\bibinfo {year}
  {2011})}\BibitemShut {NoStop}%
\bibitem [{\citenamefont {Fawcett}\ \emph {et~al.}(1991)\citenamefont
  {Fawcett}, \citenamefont {Pluzhnikov},\ and\ \citenamefont
  {Klimker}}]{CeAl2_Eric91}%
  \BibitemOpen
  \bibfield  {author} {\bibinfo {author} {\bibfnamefont {E.}~\bibnamefont
  {Fawcett}}, \bibinfo {author} {\bibfnamefont {V.}~\bibnamefont {Pluzhnikov}},
  \ and\ \bibinfo {author} {\bibfnamefont {H.}~\bibnamefont {Klimker}},\
  }\href@noop {} {\bibfield  {journal} {\bibinfo  {journal} {Phys. Rev. B}\
  }\textbf {\bibinfo {volume} {43}},\ \bibinfo {pages} {8531} (\bibinfo {year}
  {1991})}\BibitemShut {NoStop}%
\bibitem [{\citenamefont {Naito}\ \emph {et~al.}(2003)\citenamefont {Naito},
  \citenamefont {Takeuchi}, \citenamefont {Kindo}, \citenamefont {Tabata},\
  and\ \citenamefont {Kawarazaki}}]{CeRh2Si2_Naito03}%
  \BibitemOpen
  \bibfield  {author} {\bibinfo {author} {\bibfnamefont {O.}~\bibnamefont
  {Naito}}, \bibinfo {author} {\bibfnamefont {T.}~\bibnamefont {Takeuchi}},
  \bibinfo {author} {\bibfnamefont {K.}~\bibnamefont {Kindo}}, \bibinfo
  {author} {\bibfnamefont {Y.}~\bibnamefont {Tabata}}, \ and\ \bibinfo {author}
  {\bibfnamefont {S.}~\bibnamefont {Kawarazaki}},\ }\href@noop {} {\bibfield
  {journal} {\bibinfo  {journal} {Physica B: Condens. Matter}\ }\textbf
  {\bibinfo {volume} {329-333}},\ \bibinfo {pages} {512} (\bibinfo {year}
  {2003})}\BibitemShut {NoStop}%
\bibitem [{\citenamefont {Aoki}\ \emph {et~al.}(2013)\citenamefont {Aoki},
  \citenamefont {Knafo},\ and\ \citenamefont {Sheikin}}]{metamag_Aoki13}%
  \BibitemOpen
  \bibfield  {author} {\bibinfo {author} {\bibfnamefont {D.}~\bibnamefont
  {Aoki}}, \bibinfo {author} {\bibfnamefont {W.}~\bibnamefont {Knafo}}, \ and\
  \bibinfo {author} {\bibfnamefont {I.}~\bibnamefont {Sheikin}},\ }\href@noop
  {} {\bibfield  {journal} {\bibinfo  {journal} {C. R. Phys.}\ }\textbf
  {\bibinfo {volume} {14}},\ \bibinfo {pages} {53} (\bibinfo {year}
  {2013})}\BibitemShut {NoStop}%
\bibitem [{\citenamefont {Adroja}\ \emph {et~al.}()\citenamefont {Adroja},
  \citenamefont {Yang}, \citenamefont {Takabatake}, \citenamefont {Hillier},\
  and\ \citenamefont {Bhattacharyya}}]{ISIS}%
  \BibitemOpen
  \bibfield  {author} {\bibinfo {author} {\bibfnamefont {D.}~\bibnamefont
  {Adroja}}, \bibinfo {author} {\bibfnamefont {C.}~\bibnamefont {Yang}},
  \bibinfo {author} {\bibfnamefont {T.}~\bibnamefont {Takabatake}}, \bibinfo
  {author} {\bibfnamefont {A.}~\bibnamefont {Hillier}}, \ and\ \bibinfo
  {author} {\bibfnamefont {A.}~\bibnamefont {Bhattacharyya}},\ }\href@noop {}
  {\enquote {\bibinfo {title} {{G}eometrical frustration induced magnetism in
  single crystal of {CeIrSn}},}\ }\bibinfo {note} {{STFC} {ISIS} {N}eutron and
  {M}uon {S}ource, https://doi.org/10.5286/ISIS.E.RB1810852 (2018)}\BibitemShut
  {NoStop}%
\end{thebibliography}%



\maketitle
\makeatletter
\renewcommand{\thefigure}{S\@arabic\c@figure}
\makeatother
\setcounter{figure}{0}

\section*{Supplemental Material}
\section{Experimental Details}

\begin{figure}[b]
	\begin{center}
		\includegraphics[width=55mm]{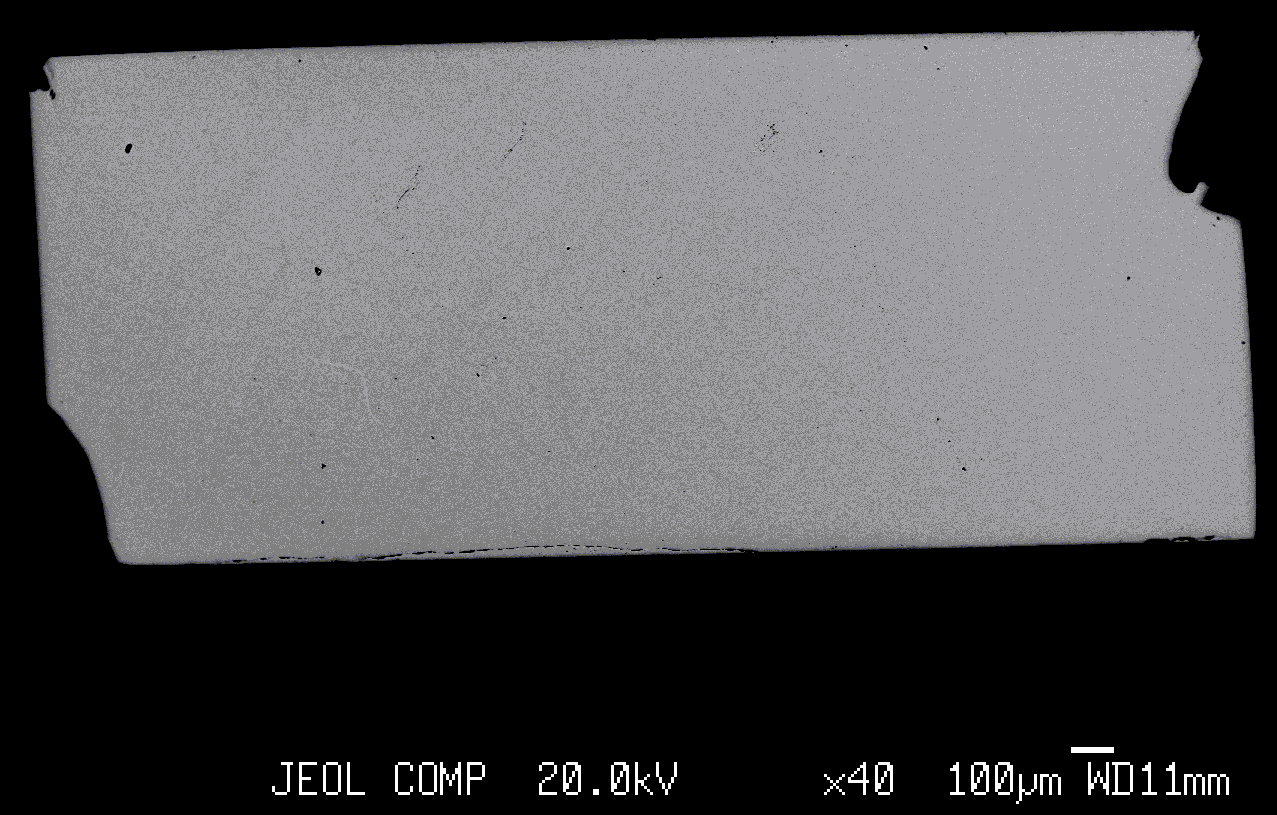}
	\end{center}
\justify
FIG.\,S1:\,\,The picture of the sample of CeIrSn used for EPMA.
\end{figure}

\begin{figure*}[t]
		\includegraphics[width=1.99\columnwidth]{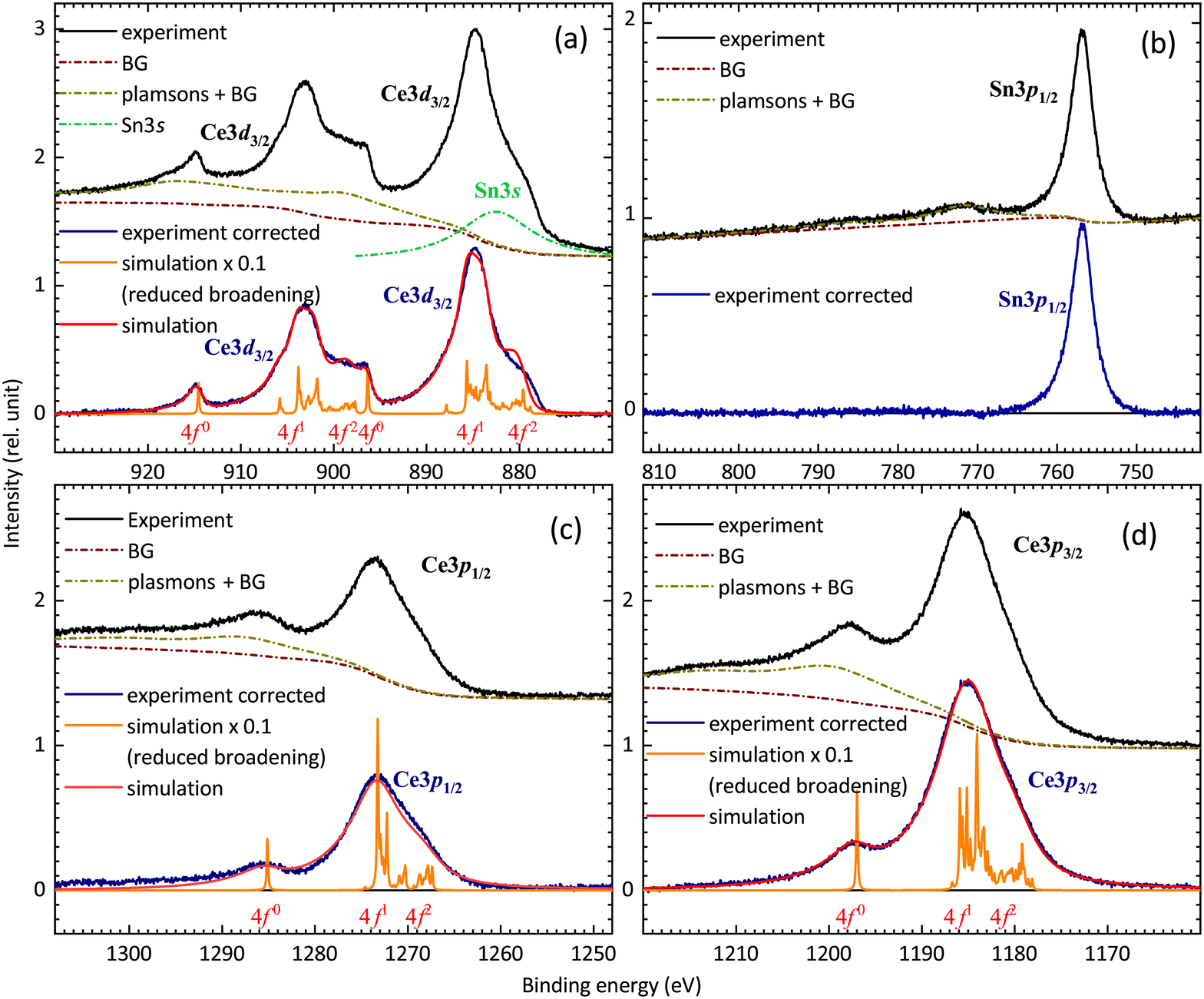} 
		\justify
FIG.\,S2:\,\,Photoemission spectra of the Ce3$d$ (a), Sn3$p_{1/2}$ (b), Ce3$p_{1/2}$ (c), and Ce3$p_{3/2}$ (d) as measured (black lines on top). The dashed lines refer to the integrated Shirley-type background (brown) and the sum of background (BG) and plamsons (dark yellow). The green line in (a) resembles the spectral weight of Sn3$s$ as obtained from the self consistent (offset by 1.2 rel.~unit). The blue lines are the Ce3$d$ and Ce3$p$ emission spectra after subtraction of background, plasmons (and Sn in case of Ce3$d$). The red lines are the result of the full-multiplet configuration-interaction simulation (fmCI). 
 \label{sup_HAXPES}
\end{figure*}

Single crystalline samples of CeIrSn were synthesized by the Czochralski method by using an RF-induction furnace \cite{CeIrSn_Tsuda18}. The atomic compositions of single crystalline samples were determined utilizing electron-probe microanalysis (EPMA) by averaging over 10 different regions.
The composition ratios for Ce : Ir : Sn = 1 : 0.97 : 1.01 were obtained by assuming that the Ce sites are fully occupied.
As shown in Fig. S1, impurity phases are not detected in our EPMA measurements.

The dilatation was measured by a miniaturized ultra-high resolution capacitive dilatometer with a width of 15.0 mm, a depth of 14.0 mm and a height of 14.7 mm \cite{Dilatometer_Kuchler17}. The compactness allows us to rotate the sample with respect to the magnetic field direction even inside the narrow space of the dilution refrigerator. It enables us to measure the dilatation on two configurations for the measured direction of the sample length parallel and perpendicular to the magnetic field. A rectangular single crystal with dimensions of 1.81\,$\times$\,1.81\,$\times$\,2.07\,mm$^3$ was used for the dilatation measurements along the hexagonal principal directions $a$, $c$, and $a^{*}$, where $a^{*}$ denotes the in-plane direction perpendicular to $a$ (cf. Fig.~S4(a)). The linear magnetostriction coefficient $\lambda $ and the linear thermal expansion coefficient $\alpha $ are defined as $\lambda _i= L_i^{-1} {\rm d} L_i/{\rm d} B$ and $\alpha _i = L_i^{-1} {\rm d} L_i/{\rm d} T$, respectively. Here, $L$ and $i$ (=$a$, $a^*$, $c$) denote the sample length and measured crystal axis, respectively. The volume magnetostriction coefficient $V^{-1}dV/dB= \lambda_V =\sum_i \lambda _i$ is obtained from the sum of the three linear magnetostriction coefficients.

The experimental procedures of the hard x-ray photo-electron spectroscopy (HAXPES) and inelastic neutron scattering are described in each section.

The $\mu$SR measurements were carried out on the EMU spectrometer at ISIS Facility, whereby the sample was cooled down to 0.1\,K by use of a dilution refrigerator. We used slices of CeIrSn single crystals with 1.5 mm thickness, which were mounted on a high-purity silver holder using GE varnish as shown in the inset of Fig. S5 (b). The analysis of the $\mu $SR data was carried out using WIMDA software \cite{Pratt00}. The crystallographic $c$-axis was aligned parallel to the muon beam and the $a$-axis vertical ($\bot $ to the muon beam). We performed zero field (ZF) and transverse field (TF) $\mu$SR measurements. The ZF data are shown in the main paper, while the TF-$\mu$SR results are presented below.

\section{HAXPES}

\begin{table}
	\caption{Parameters for the fmCI Hamiltonian. The top rows give the the resulting 4$f^n$ contributions, followed by the corresponding CI parameters for $f$-$f$ Coulomb exchange $U$($4f$,$4f$), the Coulomb interaction between the 4$f$ and 3$d$ (3$p$) core hole $U$($3d$,$4f$) ($U$($3p$,$4f$), the effective 4$f$ binding energy $\varepsilon_f$, and the hybridization strength $V_\text{eff}$. On the right the respective lineshape parameters are listed. The Lorentzian FWHM$_\text{L}$ and Gaussian width FWHM$_\text{G}$, the Mahan asymmetry factor $\alpha_M$ and cutoff energy $\gamma_M$, and the energy and width of the plasmons $\Delta$E$^\text{pl}$ and FWHM$_\text{L}^\text{pl}$.}
	\label{tab:xtlsparameters}%
	\newcommand{\mc}[2]{\multicolumn{#1}{c}{#2}}%
	\newcommand{\mcf}[3]{\multicolumn{#1}{#2}{#3}}%
	\begin{tabular*}{\columnwidth}{@{\extracolsep{\fill} } l c c || l c c c c }
	\mcf{3}{c||}{fmCI results and CI parameters} &  \mc{4}{line shape parameters}                          & \\ \hline
	c($f^0$)        & \% & 16(2)& &  & & &\\
	c($f^1$)        & \% & 81(2)& &  & & &\\
	c($f^2$)        & \% &2.7(5)&                           &    & $3d$ & $3p$  & \\
	                &    &      & FWHM$_\text{L}$           & eV &  0.4 &  4    & \\
	V$_\text{eff}$  & eV & 0.31 & FWHM$_\text{G}$           & eV & \mc{2}{1.2}  & \\
	$\varepsilon_{f}$&eV &  1.7 & $\alpha_\text{M}$         &    & \mc{2}{0.4}  & \\	
	$U$($4f$,$4f$)  & eV &  9.6 & $\gamma_\text{M}$         & eV & \mc{2}{5}    & \\
	$U$($3d$,$4f$)  & eV & 10.3 & $\Delta$E$^\text{pl}$     & eV & \mc{2}{14.5} & \\
	$U$($3p$,$4f$)  & eV & 11.0 & FWHM$_\text{L}^\text{pl}$ & eV & \mc{2}{6}    & \\
	\end{tabular*}
\end{table}

Hard x-ray photo-electron spectroscopy (HAXPES) data were taken at the P22 beamline at the PETRA-III synchrotron. The single crystalline sample was scraped under UHV condition below 10$^{-8}$\,mbar just before transfer to the measurement chamber with 5 $\times$ 10$^{-10}$\,mbar to ensure a clean surface. HAXPES data are taken at about 60\,K on the sample with an incident photon energy of 6\,keV. The Fermi edge of a Au foil was measured to convert the electron kinetic into binding energy and an instrumental resolution of about 200\,meV was measured. A scan over an wide energy range was taken to ensure absence of any emission other than from CeIrSn. The O1$s$ emission line at 530\,eV binding energy has been taken repeatedly during the experiment to ensure the surface remained clean.  

Figure\,S2\,(a) shows the deconvolution of the CeIrSn Ce3$d$ core-level spectrum (black line). The brown dashed-dotted line is the integrated Shirley-type background and the dark yellow dashed-dotted line is the sum of integrated background and plasmons. The green dashed dotted line is the estimated amount of the Sn3$s$ spectral weight. The plasmon profile was obtained from the Sn3$p$ emission line; first and second order were taken into account (see Fig.\,S2\,(b)). The Ce3$p_{1/2}$ and  Ce3$p_{3/2}$ spectra are shown in Fig.\,S2\,(c) and (d), respectively (black lines). Also here background (Shirley and some linear) and plasmons have been subtracted. The corrected spectra (blue lines at the bottom of Fig.\,S2\,(a),\,(c), and (d)) were then simulated with the combined full-multiple configuration-interaction (fmCI) calculation. The Ce3$p$ spectra are broader than the Ce3$d$ spectra but there are no other emission lines in the same energy window. Hence, they serve as a verification for the fmCI calculation describing the Ce3$d$ (red lines). 

\begin{figure}[t]
	\begin{center}
		\includegraphics[width=0.55\textwidth]{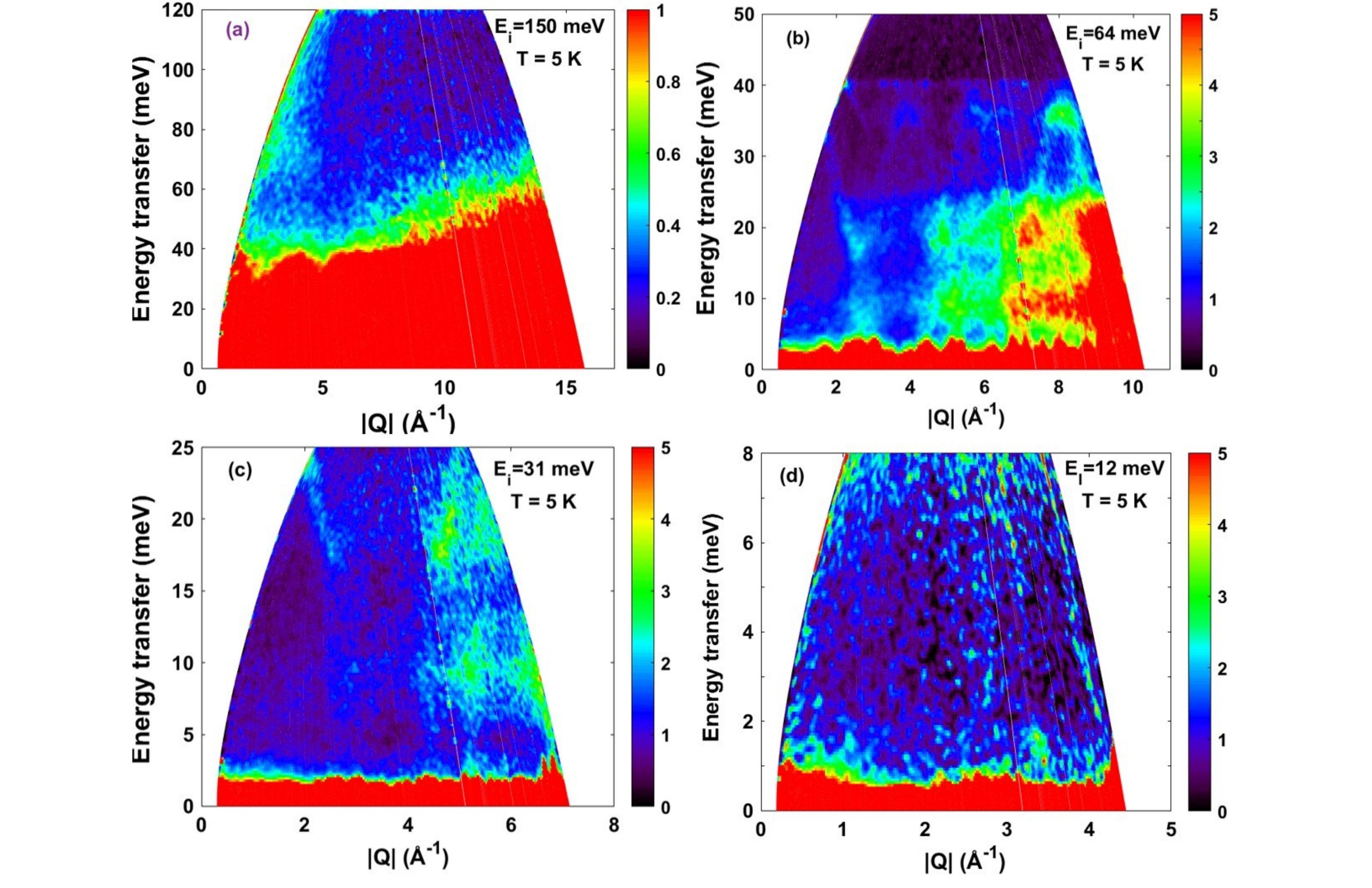}
	\end{center}
\justify
%
FIG.\,S3:\,\, Contour plots of the inelastic neutron scattering (INS) response with  energy transfer ($E$) vs. wave-vector transfer ($Q$),  of CeIrSn at 5 K, (a)  with incident energy $E_{i} = 150$ meV and chopper frequency of 500 Hz,  (b) with $E_{i} = 64$ meV and chopper frequency of 500 Hz,  (c) with $E_{i} = 31$ meV and chopper frequency of 200 Hz and (d) with $E_{i} = 12$ meV and chopper frequency of 200 Hz.  The experiments were carried out using the time of flight spectrometer MERLIN.
\end{figure}

Quantitative simulation: We applied the same method as in Refs\,\cite{Strigari2015,Sundermann2016,Sundermann2017} i.e.\ each multiplet emission line has a line profile consisting of the main emission line (Lorentzan/Gaussian plus Mahan asymmetry) and a 1$^{\rm st}$ and 2$^{\rm nd}$ order plasmon with the intensity ratios and line shapes as determined from the Sn3$p$. The line shape parameters for the Ce3$d$ and Ce3$p$ simulation are listed in Tab.\,\ref{tab:xtlsparameters}. The same line profile was then used for each emission line in the combined fmCI simulation. The simplified configuration interaction (CI) model by Imer \textit{et al.} we used here represents the valence states by only one ligand state\,\cite{Imer1987}. Its great advantage is that it can be combined with a fm calculation in a way that the computational aspects are manageable. It describes very well the Ce3$d$ and also Ce3$p$ core-level spectra but does not give realistic numbers for e.g.\ the Kondo temperature. A more detailed discussion of the advantages and short comings of the simplified model can be found in Ref.\,\cite{Strigari2015}. 

The fmCI calculations has been performed with the XTLS9.0 code by A.~Tanaka\,\cite{Tanaka1994}. The atomic values of the intra-atomic $4f$-$4f$ and $3d$-$4f$ ($3p$-$4f$) Coulomb and exchange interactions, and the 3$d$ (3$p$) and 4$f$ spin-orbit coupling are taken from the Cowan code whereby the atomic Hartree-Fock values for $3d$-$4f$ ($3p$-$4f$) Coulomb interactions are reduced to 80\%, and the spin-orbit coupling of the 3$d$ (3$p$) is reduced to 98\% (97.5\%).
The hybridization between the 4$f$ and conduction electrons is expressed in terms of the hybridization strength $V_\text{eff}$, the effective $4f$ binding energy $\varepsilon_{f}$, and the $4f$-$4f$ Coulomb exchange $U$($4f$,$4f$) in the initial state, as well as the Coulomb interaction between the 4$f$ and 3$d$ (3$p$) core hole $U$($c_i$,$4f$) with $c_i$\,=\,3$d$ (3$p$) in the final state. In total there are four parameters beside the line shape of the main emission line (Lorentzian/Gaussian and Mahan asymmetry) to be adjusted, that relate to the observed two intensity ratios and two energy distances in the photo-emission spectra. At first the energy range of the Ce3$d_{3/2}$ has been fitted and then the same parameters have been used to describe the Ce3$p_{1/2}$ and Ce$3p_{3/2}$ data. Only the Lorentzian contribution in the line shape had to be adjusted. The CI parameters in Tab.\,\ref{tab:xtlsparameters} describe both, the Ce$3d$ and Ce3$p$, very well (see red lines Fig.\,S2\,(a),\,(c), and (d)).


\section{Inelastic Neutron Scattering}

\begin{figure}[t]
	\begin{center}
		\includegraphics[width=75mm]{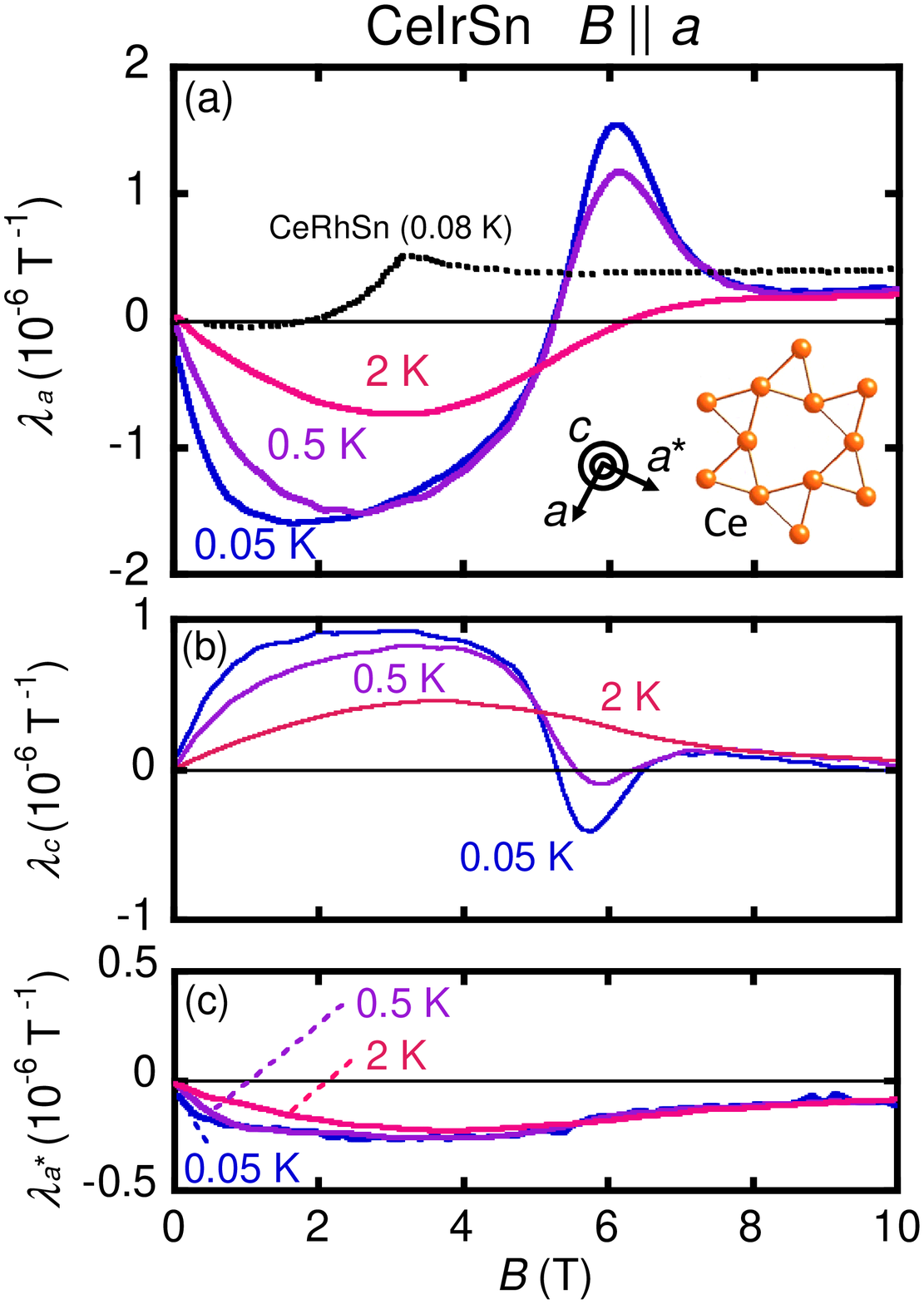}
	\end{center}
\justify
FIG.\,S4:\,\, Linear magnetostriction coefficient $\lambda _i$ for $B\parallel a$ with $i=a$ (panel a), $c$ (panel b) and $a^*$ (panel c) at different temperatures. Data for CeRhSn at 0.08\,K were extracted from Ref. \cite{CeRhSn_Kuchler17}. The three principal axes of Ce$Tr$Sn ($Tr =$ Ir and Rh) are visualized at the lower-right side in (a).
\end{figure}

We have performed inelastic neutron scattering (INS) experiments on CeIrSn at 5 K using the MERLIN spectrometer, in repetition-rate multiplication (RRM) mode that allowed us simultaneously measure with several incident energies ($E_{i}$), at the ISIS Facility, Rutherford Appleton Laboratory. We used $E_{i}$ of 150 meV and Gd-chopper frequency of 500 Hz (which also gave data for $E_{i}$=64 and 22 meV) and $E_{i}$ = 31 meV and Gd-chopper frequency of 200 Hz (which also gave data for $E_{i}$ = 12 meV). 
The Fig.S3 (a-d) shows contour maps of the inelastic neutron scattering (INS) intensity, plotted as energy transfer (E) vs wave vector transfer (Q) for $E_{i}= 150$ meV, 64 meV, 31 meV and 12 meV at 5 K. It is clear from Fig.S3 (a-d) that
there are no clear signs of well defined crystal field (CEF) excitations. This observation is in contrast with the well resolved  CEF excitations observed in Ce-based local moment compounds with trivalent Ce ions such as CeRhGe$_3$ \cite{CeRhGe3}. 
Furthermore, the scattering intensity for energy transfer below 40 meV and at high $Q$ (above $3 {\AA}^{-1}$, cf. panels (b) and (c)) indicates phononic origin.
However, the scattering intensity between  65 meV and 100 meV is stronger at lower $Q$ (cf. panel (a)), indicating magnetic origin and typically observed for strong mixed valence Ce-based compounds, i.e. CePd$_3$ \cite{CePd3_Murani96}. More detailed investigation of the magnetic INS response is left for future studies.


\section{Linear Magnetostriction for $B\parallel a$}

\begin{figure}[b]
	\begin{center}
		\includegraphics[width=65mm]{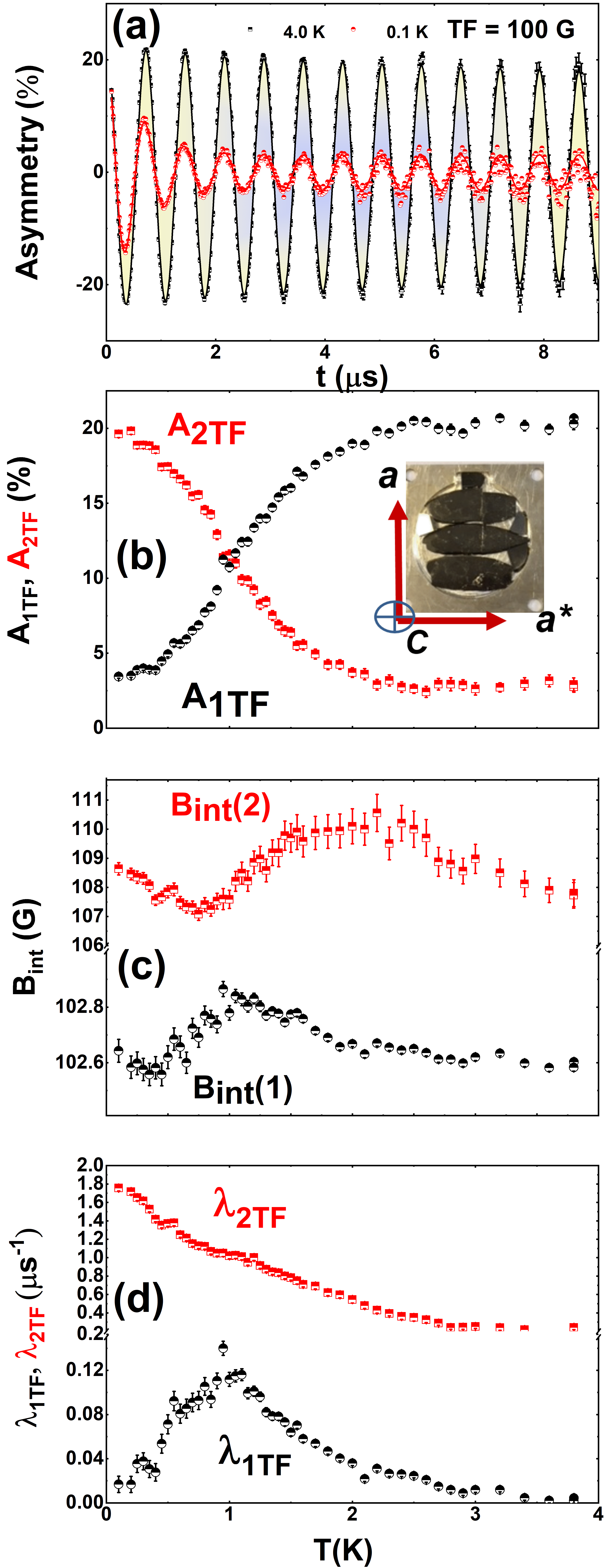}
	\end{center}
\justify
FIG.\,S5:\,\,(Color online) (a) TF-$\mu $SR (100 G parallel to the a-axis) spectra of CeIrSn at 4 K and 0.1 K. 
The solid line shows the fit to two components using Eq. (1).
Panels (b), (c) and (d) display the temperature dependence of the Asymmetries ($A_{\mathrm{1TF}}, A_{\mathrm{2TF}}$),
 internal fields ($B_{\rm int} (1), B_{\rm int} (2)$) and relaxation rates ($\lambda_{\mathrm{1TF}}, \lambda_{\mathrm{2TF}}$), respectively.
The picture in the inset of (b) shows the samples mounted on the Ag holder and the crystal axes.
\end{figure}

Figures S4 shows the field-dependence (for $B \parallel a$) of the linear magnetostriction coefficients along the three principle axes of CeIrSn at various temperatures. At 0.05\,K and 0.5\,K, $\lambda _a$ exhibits a clear peak at $B_{\rm M} = $ 6\,T which agrees with the metamagnetic crossover fields determined by magnetization and susceptibility~\cite{CeIrSn_Tsuda18}. At 2\,K, the peak is suppressed and only a broad shoulder remains. Interestingly, for fields below $B_{\rm M}$ a negative $\lambda _a$ is found which becomes more pronounced as temperature decreases. For CeRhSn (black dots in panel (a) from Ref.~\cite{CeRhSn_Kuchler17}), $\lambda _a$ also changes sign but its negative minimum at low fields is about thirty times smaller than the one of CeIrSn. The field evolution of $\lambda_c$ is reversed compared to $\lambda_a$ with a similar sign change appearing just before the sharp metamagnetic signature. By contrast, $\lambda _{a^{*}}$ varies only weakly with $B$ and displays no sign change. The reversed variation of $\lambda _a$ and $\lambda _c$ indicates that $B$ $||$ $a$ induces an anisotropic dilatation in the $a$-$c$ plane. This strongly anisotropic dilatation in the $a$-$c$ plane obscures the metamagnetic anomaly in $\lambda _{a^*}$. Mechanism of the $a$-$c$ plane coupling should be studied by neutron scattering or bulk modulus measurements in magnetic field.

\section{TF-$\mu $SR}

\begin{figure}[t]
	\begin{center}
		\includegraphics[width=80mm]{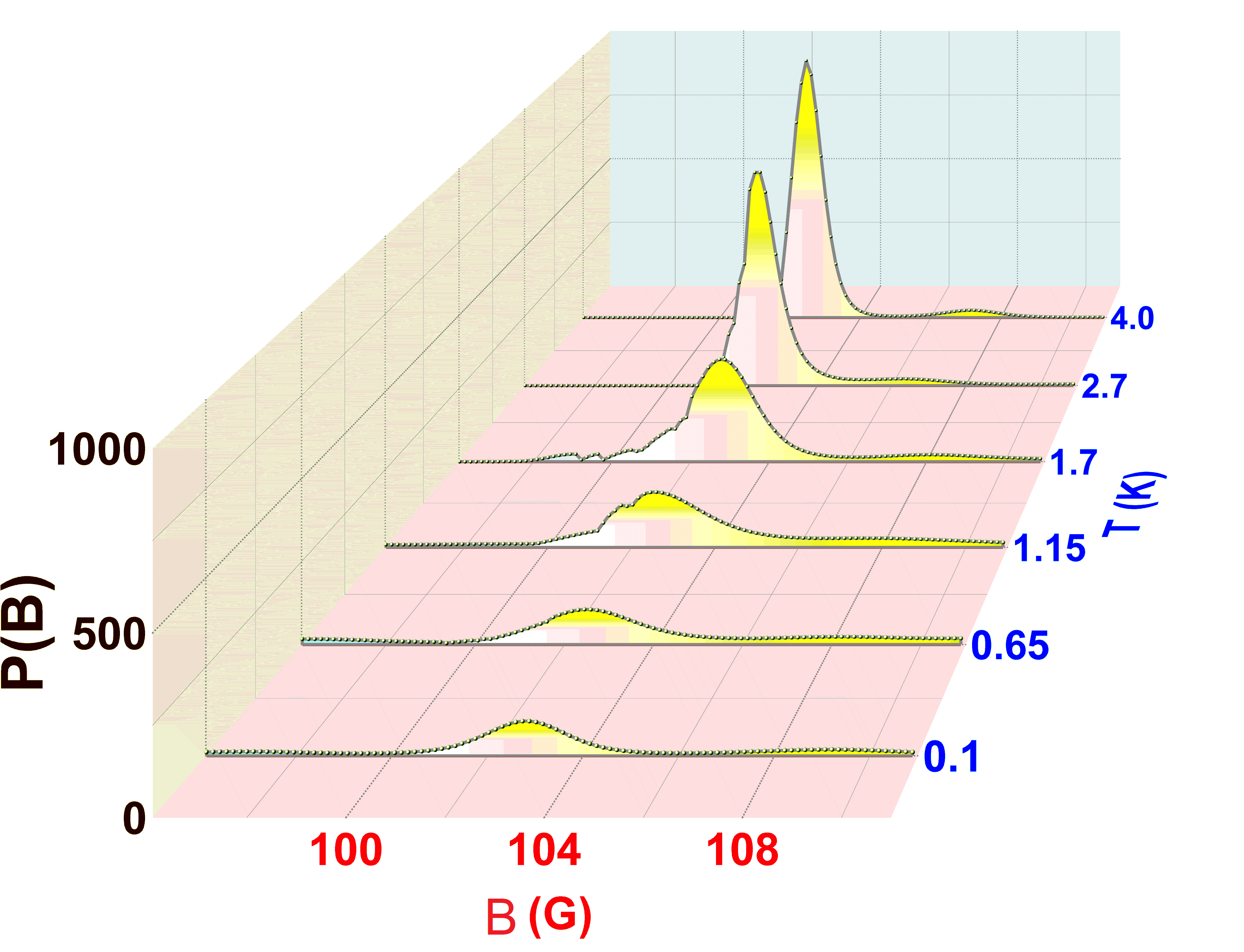}
	\end{center}
\justify
FIG.\,S6:\,\,The field distribution $P(B)$ using the maximum entropy method from the TF-$\mu $SR spectra at selected temperatures \cite{Jones91, Stephenson88, Maisuradze18}.
\end{figure}

The transverse field muon spin relaxation (TF-$\mu $SR) measurements were carried out at various temperatures from 4 K to 0.1 K with TF of 100 G applied along the a-axis (i.e., vertical direction) of the CeIrSn crystal as presented in Figs. S5 and S6. The TF-$\mu $SR spectra were fitted using the following equation
\begin{eqnarray*}
 G_z(t) &=& A_{\mathrm{1TF}} \cos (\omega_1 t + \Phi) \exp(-\lambda_{\mathrm{1TF}} t) \\
&& + A_{\mathrm{2TF}} \cos (\omega_2 t + \Phi) \exp(-\lambda_{\mathrm{2TF}} t), \ \ \ \ (1)
\end{eqnarray*}
where $A_{i\mathrm{TF}}$, $\lambda_{i\mathrm{TF}}$ and $\omega_i$ ($i =$ 1, 2), are the initial transverse-field asymmetries, Lorentzian relaxation rates, and the muon spin precession frequencies, respectively, for the slow ($i =$ 1) and the fast ($i =$ 2) components arising from the sample.
The value of the phase factor $\Phi$ was estimated by fitting the spectrum at 0.1 K, and then it was fixed for fitting spectra at all other temperatures.
The internal fields seen by the muon are given by $B_{\rm int} (i) = \gamma_\mu \omega_i$, where $\gamma_\mu$ is the muon gyromagnetic ratio.
Note that when we added the third term to account for the background from the sample holder, then the fits were not found to be reliable because of correlations among the fit parameters.

Figure S5 displays the temperature dependence of the initial asymmetries $A_{i\mathrm{TF}}$ and the relaxation rates $\lambda_{i\mathrm{TF}}$ of the slow ($i =$ 1) and fast ($i =$ 2) components obtained from the TF-$\mu $SR analysis.
Similar temperature dependences are obtained from the ZF-$\mu $SR analysis in the main text.
Furthermore, we have seen the presence of an internal field below 2 K and a peak around 1 K in the $B_{\rm int} (1)$, which are in agreement with the Maximum entropy plots given in Fig. S6. As shown in Fig. S6, the field distribution $P(B)$ near the applied field of 100 G is sharp at 4 K. 
On cooling, the peak of $P(B)$ decreases, and its width increases. 
This temperature dependence of $P(B)$ directly confirms the presence of short-range magnetic correlations below 2 K.

\end{document}